\documentclass[12pt,a4paper]{article}
\usepackage[utf8]{inputenc}
\usepackage{amsmath}
\usepackage{amsfonts} 
\usepackage{mathtools} 
\usepackage{amssymb} 
\usepackage{subfigure} 
\usepackage{graphicx}
\usepackage{xcolor}
\usepackage{cite} 
\usepackage{mathrsfs}
\usepackage{hyperref}
\usepackage{slashbox}

\graphicspath{{Figs/}}

\newcommand{\abs}[1]{|#1|} 
\newcommand{\ABS}[1]{\left|#1\right|} 
\newcommand{\re}[1]{\text{Re}\left(#1\right)}
\newcommand{\im}[1]{\text{Im}\left(#1\right)}
\newcommand{\refEQ}[1]{eq.\,\eqref{#1}} 
\newcommand{\refEQS}[1]{eqs.\,\eqref{#1}} 
\newcommand{\REFEQS}[1]{Eqs.\,\eqref{#1}} 

\newcommand{\WL}{W_{\rm L}}\newcommand{\WLd}{\WL^\dagger}
\newcommand{\WqR}[1]{W_{\rm {#1}_R}}
\newcommand{\WuR}{\WqR{u}}
\newcommand{\WdR}{\WqR{d}}
\newcommand{\UqX}[2]{\mathcal U_{#2}^{#1}}\newcommand{\UqXd}[2]{\mathcal U_{#2}^{#1\dagger}}
\newcommand{\UuL}{\UqX{u}{L}}\newcommand{\UuLd}{\UqXd{u}{L}}
\newcommand{\UdL}{\UqX{d}{L}}\newcommand{\UdLd}{\UqXd{d}{L}}
\newcommand{\UuR}{\UqX{u}{R}}
\newcommand{\UdR}{\UqX{d}{R}}
\newcommand{\Yd}[1]{\Gamma_{#1}}
\newcommand{\Yu}[1]{\Delta_{#1}}
\newcommand{\Ydd}[1]{\Gamma_{#1}^\dagger}
\newcommand{\Yud}[1]{\Delta_{#1}^\dagger}
\newcommand{\PR}[1]{{\rm P}_{\!#1}}
\newcommand{\PRX}[2]{P_{#1}^{[#2]}}
\newcommand{\PRW}[1]{\PRX{#1}{\WL}}
\newcommand{\id}{\mathbf{1}}
\newcommand{\TR}[1]{\text{Tr}\left\{#1\right\}}
\newcommand{\iTR}[1]{\im{\TR{#1}}}
\newcommand{\un}[1]{\hat n_{#1}^{\phantom{\ast}}}
\newcommand{\und}[1]{\hat n_{{\rm [d]}#1}^{\phantom{\ast}}}
\newcommand{\unu}[1]{\hat n_{{\rm [u]}#1}^{\phantom{\ast}}}
\newcommand{\unC}[1]{\hat n_{#1}^{\ast}}
\newcommand{\undC}[1]{\hat n_{{\rm [d]}#1}^{\ast}}
\newcommand{\unuC}[1]{\hat n_{{\rm [u]}#1}^{\ast}}
\newcommand{\Hd}[1]{\Phi_{#1}}
\newcommand{\Hdd}[1]{\Phi_{#1}^\dagger}
\newcommand{\Hdt}[1]{\tilde\Phi_{#1}}
\newcommand{\nHH}{\mathrm{H}^0}
\newcommand{\nHR}{\mathrm{R}^0}

\newcommand{\nh}{h}
\newcommand{\nH}{H}
\newcommand{\nA}{A}

\newcommand{\cHp}{H^+}
\newcommand{\gL}{\gamma_L}
\newcommand{\gR}{\gamma_R}
\newcommand{\cb}{c_\beta}
\renewcommand{\sb}{s_\beta}

\newcommand{\cab}{c_{\beta\alpha}}
\newcommand{\sab}{s_{\beta\alpha}}
\newcommand{\tb}{t_\beta}
\newcommand{\tbinv}{\tb^{-1}}
\newcommand{\tti}{\tb+\tbinv}

\newcommand{\CKM}{V}\newcommand{\CKMd}{\CKM^\dagger}
\newcommand{\V}[1]{{\CKM_{#1}^{\phantom{\ast}}}}
\newcommand{\Vc}[1]{{\CKM_{#1}^\ast}}

\newcommand{\mMU}{M_u}
\newcommand{\mMUd}{M_u^\dagger}
\newcommand{\wMU}{M_u^0}
\newcommand{\wMUd}{M_u^{0\dagger}}
\newcommand{\mMD}{M_d}
\newcommand{\mMDd}{M_d^\dagger}
\newcommand{\wMD}{M_d^0}
\newcommand{\wMDd}{M_d^{0\dagger}}
\newcommand{\mNQ}[1]{N_{#1}}
\newcommand{\mNU}{N_u}
\newcommand{\wNU}{N_u^{0}}
\newcommand{\mNUd}{N_u^{\dagger}}

\newcommand{\mND}{N_d}
\newcommand{\wND}{N_d^{0}}
\newcommand{\mNDd}{N_d^{\dagger}}

\newcommand{\Tvar}{\tau}
\newcommand{\ZZ}{\mathbb{Z}_{2}}
\newcounter{notas}
\graphicspath{{Figs/}}

\begin{document}
\begin{titlepage}

\hfill\begin{minipage}[r]{0.3\textwidth}\begin{flushright}  CFTP-17-04\\    IFIC-17-12 \end{flushright} \end{minipage}

\begin{center}

\vspace{0.50cm}

{\large \bf {Controlled Flavour Changing Neutral Couplings in Two Higgs Doublet Models}}

\vspace{0.50cm}

Joao M. Alves $^{a,}$\footnote{\texttt{j.magalhaes.alves@tecnico.ulisboa.pt}}
Francisco J. Botella $^{b,}$\footnote{\texttt{Francisco.J.Botella@uv.es}}, 
Gustavo C. Branco  $^{a,}$\footnote{\texttt{gbranco@tecnico.ulisboa.pt}},\\ 
Fernando Cornet-Gomez $^{b,}$\footnote{\texttt{Fernando.Cornet@ific.uv.es}}, 
Miguel Nebot $^{c,}$\footnote{\texttt{miguel.r.nebot.gomez@tecnico.ulisboa.pt}}
\end{center}

\vspace{0.50cm}
\begin{flushleft}
\emph{$^a$ Departamento de F\'\i sica and Centro de F\' \i sica Te\' orica de Part\' \i culas (CFTP),\\
\quad Instituto Superior T\' ecnico (IST), U. de Lisboa (UL),\\ 
\quad Av. Rovisco Pais, P-1049-001 Lisboa, Portugal.} \\
\emph{$^b$ Departament de F\' \i sica Te\`orica and IFIC,\\
\quad Universitat de Val\`encia-CSIC,\\
\quad E-46100, Burjassot, Spain.} \\
\emph{$^c$ Centro de F\' \i sica Te\' orica de Part\' \i culas (CFTP),\\
\quad Instituto Superior T\' ecnico (IST), U. de Lisboa (UL),\\
\quad Av. Rovisco Pais, P-1049-001 Lisboa, Portugal.} 
\end{flushleft}

\vspace{0.5cm}

\begin{abstract}
We propose a class of Two Higgs Doublet Models where there are Flavour Changing Neutral Currents (FCNC) at tree level, but under control due to the introduction of a discrete symmetry in the full Lagrangian. It is shown that in this class of models, one can have simultaneously FCNC in the up and down sectors, in contrast to the situation encountered in BGL models. The intensity of FCNC is analysed and it is shown that in this class of models one can respect all the strong constraints from experiment without unnatural fine-tuning. It is pointed out that the additional sources of flavour and CP violation are such that they can enhance significantly the generation of the Baryon Asymmetry of the Universe, with respect to the Standard Model.
\end{abstract}

\end{titlepage}

\newpage

\section{Introduction\label{SEC:INTRO}}
Recently, there has been a special interest in the scalar sector of the Standard Model (SM) and some of its extensions, motivated by the discovery by both ATLAS \cite{Aad:2012tfa} and CMS \cite{Chatrchyan:2012xdj}, of a particle which can be interpreted as the Higgs Boson of the SM. A central question is whether the Higgs couplings to quarks and leptons are like those in the SM or whether Nature chooses instead a more complex scalar sector. The simplest extension of the scalar sector consists of the addition of another Higgs doublet and the first 2 Higgs Doublet Model (2HDM) was introduced by Lee \cite{Lee:1973iz} in order to achieve spontaneous CP violation. The presence of extra symmetries has an important impact on the possibility of generating spontaneous CP violation, an aspect which has been recently reviewed \cite{Branco:2011iw,Ivanov:2017dad}. If no extra symmetries are introduced, the general 2HDM has Flavour Changing Neutral Currents (FCNC) at tree level which have
to be suppressed in order to avoid violation of the stringent experimental bounds. The simplest way of avoiding FCNC in the scalar sector is by postulating that quarks of a given charge receive contributions to their mass only from one Higgs doublet. These selective couplings \cite{Glashow:1976nt,Paschos:1976ay} can be easily implemented in a natural way  through the introduction of a $Z_2$ symmetry which leads to Natural Flavour Conservation (NFC) in the scalar sector \cite{Glashow:1976nt}. Another way of eliminating FCNC at tree level is through the hypothesis of alignment of Yukawa couplings in flavour space \cite{Pich:2009sp}. Various studies have been done on the constraints implied by FCNC, in the framework of 2HDM \cite{Mahmoudi:2009zx,Crivellin:2013wna,Gaitan:2015hga,Altunkaynak:2015twa,Arhrib:2015maa,Kim:2015zla,Enomoto:2015wbn,Benbrik:2015evd}.

A very interesting approach to the control of FCNC is provided by Branco-Grimus-Lavoura (BGL) models \cite{Branco:1996bq}, where there are FCNC at tree level but their flavour structure is controlled by the elements of the Cabibbo-Kobayashi-Maskawa (CKM) matrix $\CKM$, with no other flavour parameters. Although these models were proposed some years before the Minimal Flavour Violation (MFV) hypothesis \cite{Buras:2000dm,D'Ambrosio:2002ex,Cirigliano:2005ck} was put forward, BGL models satisfy the MFV hypothesis of having all the flavour structure of New Physics controlled by $\CKM$ \cite{Branco:1996bq,Botella:2009pq}. BGL models are renormalisable, since the flavour structure of the Yukawa couplings results from an exact symmetry of the Lagrangian. The general 2HDM contains FCNC, with their flavour structure parametrized by two complex matrices $\mND$, $\mNU$ \cite{Lavoura:1994ty}. These matrices depend on a large number of parameters, in particular on $\UdL,\UuL,\UdR$ and $\UuR$, the unitary matrices which enter in the diagonalization of the down and up quark mass matrices. Having $\mND$, $\mNU$ to depend only on $\CKM=\UuLd\UdL$ in a natural way looks like an impossible task. Yet, this task is accomplished by BGL models, which were first constructed for the quark sector and then generalized to the lepton sector \cite{Botella:2011ne}. There are six types of BGL models in the quark sector and six (three) types in the lepton sector for Dirac (Majorana) neutrinos, which
can be combined to have a total of 36 (18) BGL models, with different phenomenological implications, which were thoroughly analyzed \cite{Botella:2014ska,Bhattacharyya:2014nja,Botella:2015hoa}. An interesting feature of BGL models is the fact they contain FCNC either in the up or the down sectors but not in both sectors.

In this paper, we analyze the possibility of generalizing BGL models, having in mind the following questions:
\begin{enumerate}
\item Can one have a framework based on 2HDM which keeps the requirement of renormalisability and includes all 36 (18) models in a single model which contain each one of the BGL models as special cases?
\item Is it possible to have renormalisable 2HDM with controlled but non-vanishing FCNC in both the up and down quark sectors?
\end{enumerate}

The paper is organised as follows. In order to settle the notation, we consider in the next section the general 2HDM with no symmetry introduced, beyond the gauge symmetry. We briefly review BGL models and then propose a more general framework denoted gBGL containing BGL models in special limits. In section \ref{SEC:textures} we examine Yukawa textures corresponding to gBGL models and show how they contain BGL models as special cases.
We also derive in this section Weak Basis (WB) invariant conditions for having gBGL models. In section \ref{SEC:paramet} we propose a convenient parametrisation of gBGL models through the use of WB covariant projectors. In section \ref{SEC:scalar} we describe the scalar potential. In section \ref{SEC:fcnc} we analyse the intensity of FCNC in gBGL models, with particular emphasis, in section \ref{SEC:neartb}, on models close to BGL models of types $b$ and $t$. In section \ref{SEC:bau} the implications of gBGL models for the Baryon Asymmetry of the Universe are discussed and our conclusions are contained in the last section.

\section{Generalising BGL models: gBGL\label{SEC:gBGL}}
In order to settle the notation, we start by recalling the structure of the Yukawa couplings in the quark sector of a general 2HDM, with no extra symmetry introduced in the Lagrangian:
\begin{equation}\label{eq:Yukawa:00}
\mathscr L_{\rm Y}=
-\bar Q_L^0(\Yd{1}\Hd{1}+\Yd{2}\Hd{2}) d_R^0
-\bar Q_L^0(\Yu{1}\Hdt{1}+\Yu{2}\Hdt{2}) u_R^0+\text{h.c.}\,,
\end{equation}
with $\Hdt{j}=i\sigma_2\Hd{j}^\ast$.
Electroweak symmetry is spontaneously broken via the vacuum expectation values 
\begin{equation}\label{eq:EWSSB:00}
\langle\Hd{1}\rangle=\begin{pmatrix} 0\\ e^{i\theta_1}v_1/\sqrt{2}\end{pmatrix},\quad \langle\Hd{2}\rangle=\begin{pmatrix} 0\\ e^{i\theta_2}v_2/\sqrt{2}\end{pmatrix}.
\end{equation}
Introducing as usual $\theta=\theta_2-\theta_1$, $v^2=v_1^2+v_2^2$, $\cb=\cos\beta\equiv v_1/v$, $\sb=\sin\beta\equiv v_2/v$ and $\tb\equiv\tan\beta$, the Higgs doublets can be rotated into the ``Higgs basis'' \cite{Georgi:1978ri,Donoghue:1978cj,Botella:1994cs}
\begin{equation}\label{eq:HiggsBasis:00}
\begin{pmatrix}H_{1}\\ H_{2}\end{pmatrix}=
\begin{pmatrix}\cb & \phantom{-}\sb\\ \sb & -\cb \end{pmatrix}
\begin{pmatrix}e^{-i\theta_1}\Hd{1}\\ e^{-i\theta_2}\Hd{2}\end{pmatrix},
\end{equation}
where
\begin{equation}\label{eq:HiggsBasis:01}
\langle H_{1}\rangle=\begin{pmatrix} 0\\ v/\sqrt{2}\end{pmatrix},\quad \langle H_{2}\rangle=\begin{pmatrix} 0\\ 0\end{pmatrix}.
\end{equation}
The Yukawa couplings in \refEQ{eq:Yukawa:00} read
\begin{equation}\label{eq:Yukawa:01}
\mathscr L_{\rm Y}=
-\frac{\sqrt 2e^{i\theta_1}}{v}\bar Q_L^0(\wMD H_{1}+\wND H_{2}) d_R^0
-\frac{\sqrt 2e^{-i\theta_1}}{v}\bar Q_L^0(\wMU \tilde H_{1}+\wNU \tilde H_{2}) u_R^0+\text{h.c.}\,,
\end{equation}
with the mass matrices $\wMD$, $\wMU$, and the $\wND$, $\wNU$ matrices given by
\begin{alignat}{3}
\wMD&=\frac{v}{\sqrt 2}(\cb\Yd{1}+e^{i\theta}\sb\Yd{2})\, ,&\quad \wND&=\frac{v}{\sqrt 2}(\sb\Yd{1}-e^{i\theta}\cb\Yd{2})\, ,\label{eq:MN:matrices:00:1}\\
\wMU&=\frac{v}{\sqrt 2}(\cb\Yu{1}+e^{-i\theta}\sb\Yu{2})\, ,&\quad \wNU&=\frac{v}{\sqrt 2}(\sb\Yu{1}-e^{-i\theta}\cb\Yu{2})\, .\label{eq:MN:matrices:00:2}
\end{alignat}
It is clear from \refEQS{eq:MN:matrices:00:1}--\eqref{eq:MN:matrices:00:2} that $\wND$, $\wNU$ are complex matrices containing a huge number of new parameters, including FCNC which have strong experimental constraints. In order to avoid FCNC, Glashow and Weinberg \cite{Glashow:1976nt} introduced the NFC principle, which constrains the Yukawa couplings so that a quark of a given charge only receives mass from one Higgs doublet. BGL models also control the size of FCNC, while allowing for their appearance at tree level. They have some remarkable features which can be summarized in the following way.
\begin{enumerate}
\item[(i)] BGL models are renormalizable, since the pattern of Yukawa couplings are dictated by a symmetry of the full lagrangian.
\item[(ii)] In BGL models, the couplings of the physical neutral scalars to the quark mass eigenstates only depend on $\CKM$, $\tb$ and the quark masses, with no other parameters.
\item[(iii)] In BGL models there are FCNC either in the up or down sectors, but not in both. It has been shown \cite{Ferreira:2010ir} that if one imposes a flavour symmetry such that the FCNC only depend on $\CKM$ and further assumes that the flavour symmetry is abelian, then BGL models are unique.
\end{enumerate}
Let us recall the symmetries used in order to construct BGL models:
\begin{itemize}
\item up-type BGL models (uBGL in the following) are implemented by the symmetry
\begin{alignat}{3}\label{EQ:BGLup:00}
Q_{L_3} &\mapsto e^{i\Tvar}Q_{L_3},&\quad &\nonumber\\
d_{R} &\mapsto d_{R},&\quad &\Hd{1}\mapsto \Hd{1},&\nonumber\\
u_{R_3} &\mapsto e^{i2\Tvar}u_{R_3},&\quad &\Hd{2}\mapsto e^{i\Tvar}\Hd{2},&
\end{alignat}
\item down-type BGL models (dBGL in the following) are implemented by the symmetry
\begin{alignat}{3}\label{EQ:BGLdown:00}
Q_{L_3} &\mapsto e^{i\Tvar}Q_{L_3},&\quad &\nonumber\\
d_{R_3} &\mapsto e^{i2\Tvar}d_{R_3},&\quad &\Hd{1}\mapsto \Hd{1},&\nonumber\\
u_{R} &\mapsto u_{R},&\quad &\Hd{2}\mapsto e^{i\Tvar}\Hd{2},&
\end{alignat}
with $\Tvar\neq 0,\pi$. It has been shown \cite{Botella:2011ne} that if one extends a BGL model to the lepton sector, with Majorana neutrinos and a realistic seesaw mechanism, then $\Tvar=\pi/2$ and one is led to a $\mathbb{Z}_4$ symmetry.
\end{itemize}
In this paper, we address ourselves to the question whether it is possible to generalise BGL models so that the new class of models, called generalised BGL (gBGL), keep some of the interesting features of BGL models, like renormalizability, but allow for FCNC both in the up and the down sectors. The gBGL models are implemented through a $\ZZ$ symmetry, where $u_R$ and $d_R$ are even and only one of the scalar doublets and one of the left-handed quark doublets are odd:
\begin{alignat}{3}
Q_{L_3} &\mapsto -Q_{L_3},&\quad &\nonumber\\
d_{R} &\mapsto d_{R},&\quad &\Hd{1}\mapsto \Hd{1},&\nonumber\\
u_{R} &\mapsto u_{R},&\quad &\Hd{2}\mapsto -\Hd{2}.& \label{EQ:gBGL:00}
\end{alignat}
The above gBGL model includes all BGL models as special cases. Indeed gBGL models have some new parameters and when some of these free parameters are set to zero, one obtains a BGL model and the Lagrangian acquires a larger symmetry, namely $\mathbb{Z}_4$ \footnote{In the present work, we focus on the generalisation of BGL models in the quark sector and do not address in the detail the inclusion of the leptonic sector.}. It is worth emphasizing that gBGL models are implemented through a $\ZZ$ symmetry, as it is also the case in the Glashow--Weinberg model with NFC. The only difference is that the left-handed quark families transform differently in the two models. In words, one may say that \emph{the principle leading to gBGL constrains the Yukawa couplings so that each line of $\Yd{j}$, $\Yu{j}$ couples only to one Higgs doublet}. \\

\section{Yukawa Textures\label{SEC:textures}}
Imposing the $\ZZ$ symmetry in \refEQ{EQ:gBGL:00}, the Yukawa matrices in these models have the general form
\begin{alignat}{3}
\nonumber
&\Yd{1}=\begin{pmatrix}\times&\times&\gamma_{13}\\ \times&\times&\gamma_{23}\\ 0&0&0\end{pmatrix},&\quad
&\Yd{2}=\begin{pmatrix}0&0&0\\ 0&0&0\\ \gamma_{31}&\gamma_{32}&\times\end{pmatrix},\\
&\Yu{1}=\begin{pmatrix}\times&\times&\delta_{13}\\ \times&\times&\delta_{23}\\ 0&0&0\end{pmatrix},&\quad
&\Yu{2}=\begin{pmatrix}0&0&0\\ 0&0&0\\ \delta_{31}&\delta_{32}&\times\end{pmatrix},
\label{eq:Yukawa:Texture:00}
\end{alignat}
where $\times$, $\gamma_{ij}$ and $\delta_{ij}$ stand for arbitrary complex parameters. In \refEQ{eq:Yukawa:Texture:00}, the $\gamma_{ij}$ and $\delta_{ij}$ entries have been singled out in order to show how gBGL contain BGL models as special cases: it is evident that taking $\gamma_{ij}=0$, we obtain dBGL models, where
\begin{alignat}{3}
\nonumber
&\Yd{1}=\begin{pmatrix}\times&\times&0\\ \times&\times&0\\ 0&0&0\end{pmatrix},&\quad
&\Yd{2}=\begin{pmatrix}0&0&0\\ 0&0&0\\ 0&0&\times\end{pmatrix},\\
&\Yu{1}=\begin{pmatrix}\times&\times&\times\\ \times&\times&\times\\ 0&0&0\end{pmatrix},&\quad
&\Yu{2}=\begin{pmatrix}0&0&0\\ 0&0&0\\ \times&\times&\times\end{pmatrix},
\label{eq:Yukawa:Texture:down}
\end{alignat}
while taking $\delta_{ij}=0$ we obtain uBGL models, with
\begin{alignat}{3}
\nonumber
&\Yd{1}=\begin{pmatrix}\times&\times&\times\\ \times&\times&\times\\ 0&0&0\end{pmatrix},&\quad
&\Yd{2}=\begin{pmatrix}0&0&0\\ 0&0&0\\ \times&\times&\times\end{pmatrix},\\
&\Yu{1}=\begin{pmatrix}\times&\times&0\\ \times&\times&0\\ 0&0&0\end{pmatrix},&\quad
&\Yu{2}=\begin{pmatrix}0&0&0\\ 0&0&0\\ 0&0&\times\end{pmatrix}.
\label{eq:Yukawa:Texture:up}
\end{alignat}
That is, this class of renormalisable gBGL models includes both up and down type BGL models, as we were looking for. It is also clear that FCNC are present both in the up and in the down sectors: the appearance of FCNC in one sector depends on the Yukawa couplings of that sector alone, without regard to the Yukawa couplings in the other sector. In these gBGL models, both up and down sectors have the FCNC-inducing structure that in BGL models is confined to one and only one sector.

\subsection{Weak basis invariant conditions\label{sSEC:textures:WBI}}
As a summary of the previous discussion, gBGL models are defined by a $\ZZ$ symmetry or by the following matrix textures:
\begin{alignat}{3}
&\Yd{1}=\begin{pmatrix}\times&\times&\times\\ \times&\times&\times\\ 0&0&0\end{pmatrix},&\quad
&\Yd{2}=\begin{pmatrix}0&0&0\\ 0&0&0\\ \times&\times&\times\end{pmatrix},\nonumber\\
\label{eq:Yukawa:Texture:gBGL:00}
&\Yu{1}=\begin{pmatrix}\times&\times&\times\\ \times&\times&\times\\ 0&0&0\end{pmatrix},&\quad
&\Yu{2}=\begin{pmatrix}0&0&0\\ 0&0&0\\ \times&\times&\times\end{pmatrix}.
\end{alignat}
Obviously, these zero texture structures are valid in a particular set of Weak Basis (WB), the WB where the definition of the symmetry applies -- WB transformations are discussed in detail in section \ref{sSEC:paramet:WBIproj} --. Introducing the projection operator $\PR{3}$,
\begin{equation}\label{eq:Projector:00}
\PR{3}=\begin{pmatrix}0&0&0\\ 0&0&0\\ 0&0&1\end{pmatrix},\qquad \PR{3}\PR{3}=\PR{3},\qquad (\id-\PR{3})\PR{3}=0,
\end{equation}
it is straightforward to check that imposing the textures in \refEQS{eq:Yukawa:Texture:gBGL:00} is equivalent to the following definition of gBGL models:
\begin{alignat}{3}
&\PR{3}\Yd{1}=0,&\quad & \PR{3}\Yd{2}=\Yd{2},\nonumber\\
&\PR{3}\Yu{1}=0,&\quad & \PR{3}\Yu{2}=\Yu{2}.\label{eq:Proj:gBGL:00}
\end{alignat}
BGL models, are defined by more relations: in terms of $\PR{3}$, uBGL models satisfy
\begin{alignat}{3}
&\PR{3}\Yd{1}=0,&\quad & \PR{3}\Yd{2}=\Yd{2},\nonumber\\
&\PR{3}\Yu{1}=0,&\quad & \PR{3}\Yu{2}=\Yu{2},\label{eq:Proj:uBGL:00}\\
&\Yu{1}\PR{3}=0,&\quad & \Yu{2}\PR{3}=\Yu{2},\nonumber
\end{alignat}
while dBGL models satisfy
\begin{alignat}{3}
&\PR{3}\Yd{1}=0,&\quad & \PR{3}\Yd{2}=\Yd{2},\nonumber\\
&\PR{3}\Yu{1}=0,&\quad & \PR{3}\Yu{2}=\Yu{2},\label{eq:Proj:dBGL:00}\\
&\Yd{1}\PR{3}=0,&\quad & \Yd{2}\PR{3}=\Yd{2}.\nonumber
\end{alignat}
The last two conditions in \refEQS{eq:Proj:uBGL:00} and \eqref{eq:Proj:dBGL:00} give the block diagonal form of the Yukawa matrices in the corresponding sector (up in uBGL and down in dBGL models), enforcing the absence of FCNC in that sector.
From these conditions valid in a set of WB, we can get WB independent matrix conditions for all three types of models. The conditions of interest for gBGL models are
\begin{alignat}{3}
&\Ydd{2}\Yd{1}=0,&\quad & \Ydd{2}\Yu{1}=0,\nonumber\\
&\Yud{2}\Yu{1}=0,&\quad & \Yud{2}\Yd{1}=0.\label{eq:gBGLcond:00}
\end{alignat}
Notice that \refEQS{eq:gBGLcond:00} are satisfied trivially in case $\Yd{1}=\Yu{1}=0$ or in case $\Yd{2}=\Yu{2}=0$, which correspond to 2HDM of types I or X;
note, however, that this kind of models are not of the gBGL type. Coming back to \refEQS{eq:gBGLcond:00}, it is straightforward to show that they are necessary conditions for gBGL models since, from \refEQ{eq:Proj:gBGL:00},
\begin{equation}
\Ydd{2}\Yd{1}=(\PR{3}\Yd{2})^\dagger\Yd{1}=(\Ydd{2}\PR{3})\Yd{1}=\Ydd{2}(\PR{3}\Yd{1})=0,
\end{equation}
and similarly for the remaining conditions in \refEQ{eq:gBGLcond:00}. The sufficiency of these conditions in order to have gBGL models is shown in appendix \ref{SEC:APP:cond}, where the relation with 2HDM of type I is also analysed.

\section{Parametrisation of gBGL models\label{SEC:paramet}}
It is clear that gBGL models have a great reduction in the number of free parameters, with respect to the general 2HDM. In this section, we use projection operators to suggest some convenient parametrisations of gBGL models.
\subsection{Weak basis invariant projectors\label{sSEC:paramet:WBIproj}}
Let us recall that under a WB transformation we have
\begin{equation}\label{eq:WB:fields:00}
Q^0_L\mapsto Q_L^{0\prime}=\WL Q^0_L\,;\ d^0_R\mapsto d_R^{0\prime}=\WdR d^0_R\,;\ u^0_R\mapsto u_R^{0\prime}=\WuR u^0_R,
\end{equation}
and
\begin{equation}\label{eq:WB:yukawas:00}
\Yd{i} \mapsto \Yd{i}^\prime=\WLd \Yd{i}\WdR\,;\quad \Yu{i} \mapsto \Yu{i}^\prime=\WLd \Yu{i}\WuR,
\end{equation}
with $\WL$, $\WdR$ and $\WuR$ unitary matrices.\\ 
If we now take the gBGL definition through projectors in \refEQ{eq:Proj:gBGL:00} for $\Yd{i}$ and $\Yu{i}$, and go to an arbitrary weak basis, we have
\begin{alignat}{3}
&\WLd\PR{3}\WL\WLd\Yd{1}\WdR=0,&\quad & \WLd\PR{3}\WL\WLd\Yd{2}\WdR=\WLd\Yd{2}\WdR,\nonumber\\
&\WLd\PR{3}\WL\WLd\Yu{1}\WuR=0,&\quad & \WLd\PR{3}\WL\WLd\Yu{2}\WuR=\WLd\Yu{2}\WuR.\label{eq:WB:Proj:00}
\end{alignat}
These equations are valid for any weak basis. Introducing the projector
\begin{equation}\label{eq:WB:Proj:01}
\PRW{3}=\WLd\PR{3}\WL,
\end{equation}
we have
\begin{alignat}{3}
&\PRW{3}\Yd{1}^\prime=0,&\quad & \PRW{3}\Yd{2}^\prime=\Yd{2}^\prime,\nonumber\\
&\PRW{3}\Yu{1}^\prime=0,&\quad & \PRW{3}\Yu{2}^\prime=\Yu{2}^\prime,\label{eq:WB:Proj:02}
\end{alignat}
for an arbitrary WB (from now on, we drop the primes). If we choose, for convenience,
\begin{equation}\label{eq:WB:Proj:up:00}
\WLd=\UuLd\mathcal U,
\end{equation}
where $\UuLd$ appears in the usual bi-diagonalisation $\UuLd \wMU \UuR=\mMU$ (see section \ref{sSEC:paramet:Mass}), then $\mathcal U$ is an arbitrary unitary matrix and we have a general WB invariant parametrisation in terms of that arbitrary $\mathcal U$ and of $\UuLd$; \refEQS{eq:WB:Proj:01} and \eqref{eq:WB:Proj:02} read
\begin{equation}\label{eq:WB:Proj:up:01}
\PRX{\mathcal U 3}{u_L}=\UuL(\mathcal U\PR{3}\mathcal U^\dagger)\UuLd,
\end{equation}
and
\begin{alignat}{3}
&\PRX{\mathcal U 3}{u_L}\Yd{1}=0,&\quad & \PRX{\mathcal U 3}{u_L}\Yd{2}=\Yd{2},\nonumber\\
&\PRX{\mathcal U 3}{u_L}\Yu{1}=0,&\quad & \PRX{\mathcal U 3}{u_L}\Yu{2}=\Yu{2}.\label{eq:WB:Proj:up:02}
\end{alignat}
Notice that this $\UuL$ dependence reminds us of uBGL models. An alternative parametrisation is obtained taking
\begin{equation}\label{eq:WB:Proj:down:00}
\WLd=\UdL\mathcal U^\prime,
\end{equation}
where in this case $\UdLd$ comes from the down mass matrix diagonalisation $\UdLd \wMD \UdR=\mMD$, with
\begin{equation}\label{eq:WB:Proj:down:01}
\PRX{\mathcal U^\prime 3}{d_L}=\UdL(\mathcal U^\prime\PR{3}\mathcal U^{\prime\dagger})\UdLd,
\end{equation}
and
\begin{alignat}{3}
&\PRX{\mathcal U^\prime 3}{d_L}\Yd{1}=0,&\quad & \PRX{\mathcal U^\prime 3}{d_L}\Yd{2}=\Yd{2},\\
&\PRX{\mathcal U^\prime 3}{d_L}\Yu{1}=0,&\quad & \PRX{\mathcal U^\prime 3}{d_L}\Yu{2}=\Yu{2},
\end{alignat}
where now the $\UdL$ dependence reminds us of dBGL models. Identifying \refEQS{eq:WB:Proj:up:00} and \eqref{eq:WB:Proj:down:00},
\begin{equation}
\WLd=\UuL\mathcal U=\UdL\mathcal U^\prime\,,
\end{equation}
and it is straightforward to conclude that one can use equivalently one or the other parametrisation of the same model provided 
\begin{equation}
\CKMd=\UdLd\UuL=\mathcal U^\prime\mathcal U^\dagger,\quad \text{i.e. }\mathcal U\mathcal U^{\prime\dagger}=\CKM.
\end{equation}

\subsection{Weak basis covariant parametrisation\label{sSEC:paramet:WBcovpar}}
We have in the general 2HDM
\begin{align}
\wND&=\tb\wMD-(\tti)\frac{v}{\sqrt 2}e^{i\theta}\Yd{2}\, ,\label{eq:N:matrices:00:1}\\
\wNU&=\tb\wMU-(\tti)\frac{v}{\sqrt 2}e^{-i\theta}\Yu{2}\, .\label{eq:N:matrices:00:2}
\end{align}
Assuming the existence of the projection operator $\PRX{X 3}{q_L}$ satisfying
\begin{alignat}{3}
&\PRX{X 3}{q_L}\Yd{1}=0,&\quad & \PRX{X 3}{q_L}\Yd{2}=\Yd{2}\,,\label{eq:ProjX3:00:1}\\
&\PRX{X 3}{q_L}\Yu{1}=0,&\quad & \PRX{X 3}{q_L}\Yu{2}=\Yu{2}\,,\label{eq:ProjX3:00:2}
\end{alignat}
with $X$ and $\PRX{X 3}{q_L}$ to be specified later,
\begin{align}
\PRX{X 3}{q_L}\wMD&=\PRX{X 3}{q_L}\frac{v}{\sqrt 2}(\cb\Yd{1}+e^{i\theta}\sb\Yd{2})=\frac{ve^{i\theta}}{\sqrt 2}\sb\PRX{X 3}{q_L}\Yd{2}=\frac{ve^{i\theta}}{\sqrt 2}\sb\Yd{2}\, ,\label{eq:ProjX3:01:1}\\
\PRX{X 3}{q_L}\wMU&=\PRX{X 3}{q_L}\frac{v}{\sqrt 2}(\cb\Yu{1}+e^{-i\theta}\sb\Yu{2})=\frac{ve^{-i\theta}}{\sqrt 2}\sb\PRX{X 3}{q_L}\Yu{2}=\frac{ve^{-i\theta}}{\sqrt 2}\sb\Yu{2}\, .\label{eq:ProjX3:01:2}
\end{align}
Equations \eqref{eq:N:matrices:00:1} and \eqref{eq:N:matrices:00:2} can then be rewritten as the general WB covariant gBGL parametrisation
\begin{align}
\wND&=\left[\tb\id-(\tti)\PRX{X 3}{q_L}\right]\wMD\, ,\label{eq:N:ProjX3:00:1}\\
\wNU&=\left[\tb\id-(\tti)\PRX{X 3}{q_L}\right]\wMU\, .\label{eq:N:ProjX3:00:2}
\end{align}
We can choose the up parametrisation for $\PRX{X 3}{q_L}$,
\begin{equation}\label{eq:ProjX3:up:00}
\PRX{X 3}{q_L}=\PRX{\mathcal U 3}{u_L}=\UuL(\mathcal U\PR{3}\mathcal U^\dagger)\UuLd\,,
\end{equation}
or, equivalently, we can choose the down parametrisation for $\PRX{X 3}{q_L}$,
\begin{equation}\label{eq:ProjX3:down:00}
\PRX{X 3}{q_L}=\PRX{\mathcal U^\prime 3}{d_L}=\UdL(\mathcal U^\prime\PR{3}\mathcal U^{\prime\dagger})\UdLd\,,
\end{equation}
with $\CKM=\mathcal U\mathcal U^{\prime\dagger}$, to completely define the model. Let us analyse the details of these parametrisations, after we finally rotate the quark fields to the mass basis. 

\subsection{Parametrisations in the quark mass basis\label{sSEC:paramet:Mass}}
\begin{figure}
\begin{center}
\subfigure[Down parametrisation.\label{fig:gBGLmodels:down}]{\includegraphics[width=0.35\textwidth]{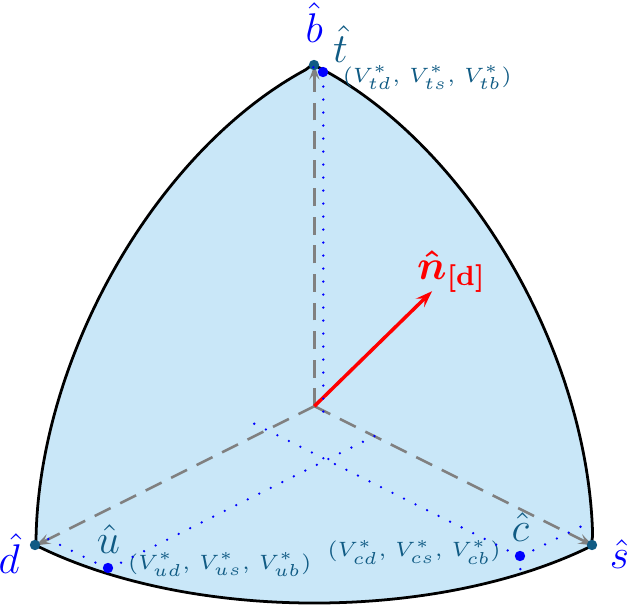}}\qquad
\subfigure[Up parametrisation.\label{fig:gBGLmodels:up}]{\includegraphics[width=0.35\textwidth]{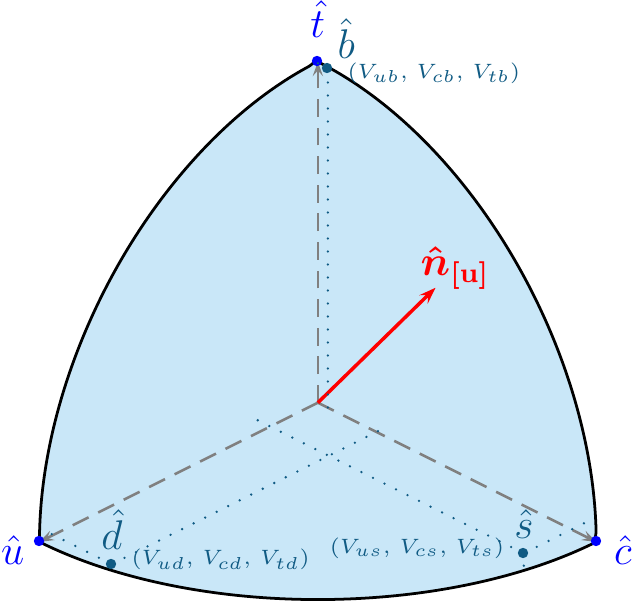}}\\
\subfigure[Generic parametrisation.\label{fig:gBGLmodels:gen}]{\includegraphics[width=0.55\textwidth]{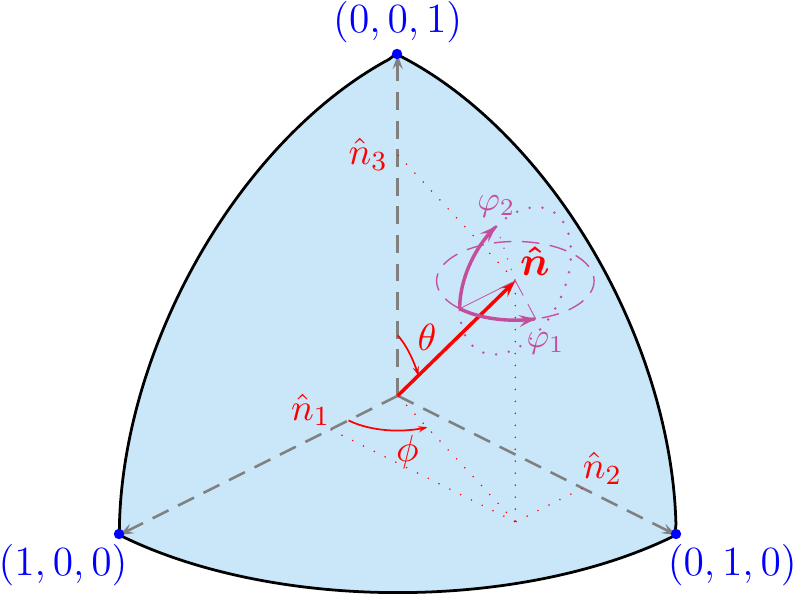}}
\caption{gBGL models.\label{fig:gBGLmodels}}
\end{center}
\end{figure}

Quark fields are rotated in the following manner
\begin{equation}\label{eq:MassBasis:00}
u_L^0=\UuL u_L,\ u_R^0=\UuR u_R,\ d_L^0=\UdL d_L,\ d_R^0=\UdR d_R,
\end{equation}
with unitary $\UqX{q}{X}$ ($q=u,d$, $X=L,R$), such that
\begin{equation}\label{eq:MassBasis:01}
\UuLd\wMU\UuR=\mMU=\text{diag}(m_u,m_c,m_t),\quad \UdLd\wMD\UdR=\mMD=\text{diag}(m_d,m_s,m_b).
\end{equation}
$\wNU$ and $\wND$ are transformed accordingly,
\begin{equation}\label{eq:MassBasis:02}
\wNU\mapsto \mNU=\UuLd\wNU\UuR,\quad \wND\mapsto \mND=\UdLd\wND\UdR\,,
\end{equation}
and, following \refEQS{eq:N:ProjX3:00:1}--\eqref{eq:N:ProjX3:00:2},
\begin{align}
\mND&=\left[\tb\id-(\tti)\UdLd\PRX{X 3}{q_L}\UdL\right]\mMD\, ,\label{eq:N:ProjX3:01:1}\\
\mNU&=\left[\tb\id-(\tti)\UuLd\PRX{X 3}{q_L}\UuL\right]\mMU\, .\label{eq:N:ProjX3:01:2}
\end{align}
If we choose the down parametrisation in \refEQ{eq:ProjX3:down:00}, \refEQS{eq:N:ProjX3:01:1}--\eqref{eq:N:ProjX3:01:2} give
\begin{align}
\mND&=\left[\tb\id-(\tti)\mathcal U^\prime\PR{3}\mathcal U^{\prime\dagger}\right]\mMD\, ,\label{eq:N:Proj3down:00:1}\\
\mNU&=\left[\tb\id-(\tti)\CKM\mathcal U^\prime\PR{3}\mathcal U^{\prime\dagger}\CKMd\right]\mMU\, ,\label{eq:N:Proj3down:00:2}
\end{align}
while, if we choose the up parametrisation in \refEQ{eq:ProjX3:up:00}, \refEQS{eq:N:ProjX3:01:1}--\eqref{eq:N:ProjX3:01:2} give
\begin{align}
\mND&=\left[\tb\id-(\tti)\CKMd\mathcal U\PR{3}\mathcal U^{\dagger}\CKM\right]\mMD\, ,\label{eq:N:Proj3up:00:1}\\
\mNU&=\left[\tb\id-(\tti)\mathcal U\PR{3}\mathcal U^{\dagger}\right]\mMU\, .\label{eq:N:Proj3up:00:2}
\end{align}
Equations \eqref{eq:N:Proj3down:00:1}--\eqref{eq:N:Proj3down:00:2} and \eqref{eq:N:Proj3up:00:1}--\eqref{eq:N:Proj3up:00:2} are of course equivalent, since $\CKM=\mathcal U\mathcal U^{\prime\dagger}$.
At this point it is important to discuss a central question: it may appear that there is a large arbitrariness in the definition of the model since there is a completely arbitrary unitary matrix $\mathcal U$ involved in \refEQS{eq:N:Proj3up:00:1}--\eqref{eq:N:Proj3up:00:2}. Nevertheless, there is much less freedom since the quantities involving $\mathcal U$ are 
\begin{equation}\label{eq:UP3Udag:00}
[\mathcal U\PR{3}\mathcal U^\dagger]_{ij}=\mathcal U_{i3}^{\phantom{\ast}}\mathcal U_{j3}^{\ast},
\end{equation}
that is, only the elements of the third column of $\mathcal U$, which form a unitary complex vector, are needed to define the model. To stress this fact, we introduce
\begin{equation}\label{eq:n:vec:00}
\unu{i}\equiv\mathcal U_{i3},\quad\text{and}\quad \und{i}=\mathcal U^\prime_{i3},\quad\text{with}\quad \unu{j}=\V{ji}\und{i}\,,
\end{equation}
where the subindex $[u]$ or $[d]$ specifies the parametrisation under consideration. The matrix elements of $\mNU$ and $\mND$ in equations \eqref{eq:N:ProjX3:01:1} and \eqref{eq:N:ProjX3:01:2} can be written, explicitely,
\begin{align}
[\mND]_{ij}&=\tb\delta_{ij}m_{d_i}-(\tti)\und{i}\undC{j}m_{d_j}\, ,\label{eq:Nq:gen:00:1}\\
[\mNU]_{ij}&=\tb\delta_{ij}m_{u_i}-(\tti)\unu{i}\unuC{j}m_{u_j}\, .\label{eq:Nq:gen:00:2}
\end{align}
Since $\unu{j}=\V{ji}\und{i}$, one is free to rewrite \refEQS{eq:N:Proj3down:00:1}--\eqref{eq:N:Proj3down:00:2} as
\begin{align}
[\mND]_{ij}&=\tb\delta_{ij}m_{d_i}-(\tti)\und{i}\undC{j}m_{d_j}\, ,\label{eq:Nq:down:00:1}\\
[\mNU]_{ij}&=\tb\delta_{ij}m_{u_i}-(\tti)\V{ia}\Vc{jb}\und{a}\undC{b}m_{u_j}\, .\label{eq:Nq:down:00:2}
\end{align}
or \refEQS{eq:N:Proj3up:00:1}--\eqref{eq:N:Proj3up:00:2} as
\begin{align}
[\mND]_{ij}&=\tb\delta_{ij}m_{d_i}-(\tti)\unu{a}\unuC{b}\Vc{ai}\V{bj}m_{d_j}\, ,\label{eq:Nq:up:00:1}\\
[\mNU]_{ij}&=\tb\delta_{ij}m_{u_i}-(\tti)\unu{i}\unuC{j}m_{u_j}\, .\label{eq:Nq:up:00:2}
\end{align}
One can now identify easily all the usual BGL models with this parametrisation. The notation is straightforward, for example, BGL model ``$s$'' corresponds to $\hat s$, and so on:
\begin{itemize}
\item dBGL models in the down parametrisation,
\begin{equation}\label{eq:dBGL:down}
\hat d_{\rm [d]}=\begin{pmatrix}1\\ 0\\ 0\end{pmatrix},\quad
\hat s_{\rm [d]}=\begin{pmatrix}0\\ 1\\ 0\end{pmatrix},\quad
\hat b_{\rm [d]}=\begin{pmatrix}0\\ 0\\ 1\end{pmatrix},
\end{equation}
\item uBGL models in the up parametrisation,
\begin{equation}\label{eq:uBGL:up}
\hat u_{\rm [u]}=\begin{pmatrix}1\\ 0\\ 0\end{pmatrix},\quad
\hat c_{\rm [u]}=\begin{pmatrix}0\\ 1\\ 0\end{pmatrix},\quad
\hat t_{\rm [u]}=\begin{pmatrix}0\\ 0\\ 1\end{pmatrix},
\end{equation}
\item dBGL models in the up parametrisation,
\begin{equation}\label{eq:dBGL:up}
\hat d_{\rm [u]}=\begin{pmatrix}\V{ud}\\ \V{cd}\\ \V{td}\end{pmatrix},\quad
\hat s_{\rm [u]}=\begin{pmatrix}\V{us}\\ \V{cs}\\ \V{ts}\end{pmatrix},\quad
\hat b_{\rm [u]}=\begin{pmatrix}\V{ub}\\ \V{cb}\\ \V{tb}\end{pmatrix},
\end{equation}
\item uBGL models in the down parametrisation,
\begin{equation}\label{eq:uBGL:down}
\hat u_{\rm [d]}=\begin{pmatrix}\Vc{ud}\\ \Vc{us}\\ \Vc{ub}\end{pmatrix},\quad
\hat c_{\rm [d]}=\begin{pmatrix}\Vc{cd}\\ \Vc{cs}\\ \Vc{cb}\end{pmatrix},\quad
\hat t_{\rm [d]}=\begin{pmatrix}\Vc{td}\\ \Vc{ts}\\ \Vc{tb}\end{pmatrix}.
\end{equation}
\end{itemize}
It is also possible to give a graphical description of the gBGL class of models, as shown in figure \ref{fig:gBGLmodels}. Since a model is defined by a complex unitary vector $\hat n$, $\abs{\un{1}}^2+\abs{\un{2}}^2+\abs{\un{3}}^2=1$, and $(\abs{\un{1}},\abs{\un{2}},\abs{\un{3}})$ is located on the sphere of unit radius (specifically, on an octant of that sphere). Furthermore, there are two physical complex phases, since one can readily see that $\un{i}\unC{j}$ is unaffected by a global rephasing of $\hat n$, which can be used to remove one out of the three initial phases in the $\hat n$ components. This is illustrated in figure \ref{fig:gBGLmodels:gen}, where no explicit reference to up, down or other parametrisations is made\footnote{Notice that one can trivially adopt spherical coordinates with, for example,  $(\abs{\un{1}},\abs{\un{2}},\abs{\un{3}})=(\sin\theta\cos\phi,\sin\theta\sin\phi,\cos\theta)$.}. Figures \ref{fig:gBGLmodels:down} and \ref{fig:gBGLmodels:up} illustrate the situation for down and up parametrisations, including the usual BGL models in \refEQS{eq:dBGL:down} to \eqref{eq:uBGL:down}.

\section{The scalar sector\label{SEC:scalar}}
The scalar potential, imposing invariance under $\Hd{2}\mapsto -\Hd{2}$ in \refEQ{EQ:gBGL:00}, is
\begin{multline}\label{eq:ScalarPotential:00}
V(\Hd{1},\Hd{2})=
\mu_{11}^2\Hdd{1}\Hd{1}+\mu_{22}^2\Hdd{2}\Hd{2}+\lambda_1(\Hdd{1}\Hd{1})^2+\lambda_2(\Hdd{2}\Hd{2})^2\\
+2\lambda_3(\Hdd{1}\Hd{1})(\Hdd{2}\Hd{2})+2\lambda_4(\Hdd{1}\Hd{2})(\Hdd{2}\Hd{1})+[\lambda_5(\Hdd{1}\Hd{2})^2+\text{h.c.}]\,.
\end{multline}
Obviously, this scalar potential coincides with the one in Glashow-Weinberg model \cite{Glashow:1976nt}, since in both cases a $\ZZ$ symmetry is introduced. In BGL models where instead of $\ZZ$ one uses a larger symmetry, namely $\mathbb{Z}_4$, the term in $\lambda_5$ is not $\mathbb{Z}_4$ invariant and therefore cannot be introduced. This leads to a global symmetry which upon spontaneous breaking would lead to a Golstone boson. This difficulty can be avoided by softly breaking the discrete symmetry through the addition of a term $(m_{12}^2\Hdd{1}\Hd{2}+\text{h.c.})$. The scalar potential of \refEQ{eq:ScalarPotential:00} does not lead to new sources of CP violation. But when the above bilinear term is introduced, one can have either spontaneous \cite{Branco:1985aq} or explicit CP breaking in the scalar sector. 
Expanding 
\begin{equation}\label{eq:DoubletPhi:00}
\Hd{j}=e^{i\theta_j}\begin{pmatrix}\varphi_j^+\\ (v_j+\rho_j+i\eta_j)/\sqrt{2}\end{pmatrix},
\end{equation}
then the rotation to the Higgs basis in \refEQ{eq:HiggsBasis:00} identifies the Goldstone boson interpretation of $G^\pm$ and $G^0$ as the longitudinal degrees of freedom of the $W^\pm$ and $Z^0$ gauge bosons,
\begin{equation}\label{eq:HiggsBasis:02}
H_1=\begin{pmatrix}G^+\\ (v+\nHH+iG^0)/\sqrt{2}\end{pmatrix}\,,\qquad
H_2=\begin{pmatrix}\cHp\\ (\nHR+i\nA)/\sqrt{2}\end{pmatrix}\,,
\end{equation}
and
\begin{equation}\label{eq:NeutralScalars:00}
\begin{pmatrix}\nHH\\ \nHR\end{pmatrix} = \begin{pmatrix}\cb & \sb\\ \sb & -\cb\end{pmatrix}\begin{pmatrix}\rho_1\\ \rho_2\end{pmatrix}\,.
\end{equation}
The scalar potential in \refEQ{eq:ScalarPotential:00} is in general CP conserving in such a way that  the physical CP even states do not mix with the CP odd one; then, the physical scalars are a linear combination of $\rho_1$ and $\rho_2$,
\begin{equation}\label{eq:NeutralScalars:01}
\begin{pmatrix}\nH\\ \nh\end{pmatrix}=\begin{pmatrix}\cos\alpha & \sin\alpha\\ -\sin\alpha & \cos\alpha\end{pmatrix}\begin{pmatrix}\rho_1\\ \rho_2\end{pmatrix}\,,
\end{equation}
and we will have in general
\begin{equation}\label{eq:NeutralScalars:02}
\begin{pmatrix}\nHH\\ \nHR\end{pmatrix}=\begin{pmatrix}\cab & \sab\\ \sab & -\cab\end{pmatrix}\begin{pmatrix}\nH\\ \nh\end{pmatrix}\,,
\end{equation}
where the relevant angle is $\beta-\alpha$ and $\cab\equiv\cos(\beta-\alpha)$, $\sab\equiv\sin(\beta-\alpha)$. The field $\nA$ is a physical pseudoscalar. Notice that the Yukawa couplings become
\begin{align}
\mathscr L_{\rm Y}= &-\frac{\sqrt 2}{v}\cHp\bar u(\CKM\mND\gR-\mNUd\CKM\gL)d+\text{h.c.}\nonumber\\
&-\frac{1}{v}\nHH(\bar u\mMU u+\bar d\mMD d)\nonumber\\
&-\frac{1}{v}\nHR\left[\bar u(\mNU\gR+\mNUd\gL) u + \bar d(\mND\gR+\mNDd\gL)d\right]\nonumber\\
&+\frac{i}{v}\nA\left[\bar u(\mNU\gR-\mNUd\gL) u - \bar d(\mND\gR-\mNDd\gL)d\right]\,,\label{eq:YukawasFull:00}
\end{align}
in such a way that, for $\nh\to\nHH$, $\nh$ becomes ``the standard Higgs'', coupling to fermions like the SM Higgs does. That is, when $\sab=1$ we have $\nh$ aligned with $\nHH$ with SM couplings.
 
\section{The intensity of FCNC in gBGL\label{SEC:fcnc}}
FCNC are extremely constrained by experimental data, consequently we have to worry about their intensity in gBGL models. It is worthwhile to mention that BGL are Minimal Flavour Violating (MFV) models, meaning that FCNC are controlled by the deviation of the CKM matrix $\CKM$ from $\id$. As a result, in the limit $\CKM=\id$, there are no FCNC in BGL models at tree level. However, this is not the case in gBGL models, where FCNC are no longer controlled by the deviations of $\CKM$ from unity. The Yukawa couplings of gBGL models are given, following \refEQS{eq:YukawasFull:00} and \eqref{eq:NeutralScalars:02}, by
\begin{equation}
\mathscr L_{\rm h\bar qq}=-\sum_{i,j=d,s,b}\nh\bar d_{L_i} Y^{\rm D}_{ij}d_{R_j}-\sum_{i,j=u,c,t}\nh\bar u_{L_i} Y^{\rm U}_{ij}u_{R_j}\,,
\end{equation}
with
\begin{align}
Y^{\rm D}_{ij}&=\frac{1}{v}\left[\sab(\mMD)_{ij}+\cab(\mND)_{ij}\right]\,,\nonumber\\
Y^{\rm U}_{ij}&=\frac{1}{v}\left[\sab(\mMU)_{ij}+\cab(\mNU)_{ij}\right]\,,
\end{align}
with $\mNQ{q}$ given in \refEQS{eq:Nq:gen:00:1}--\eqref{eq:Nq:gen:00:2}, that generically can be written
\begin{equation}
(\mNQ{q})_{ij}=\left[\tb\delta_{ij}-(\tti)\un{{\rm [q]}i}\unC{{\rm [q]}j}\right]m_{q_j}\,.
\end{equation}
So, the flavour changing intensities are controlled by $\mNQ{q}$ with the following factors.
\begin{itemize}
\item In an $q_{R_j}\to q_{L_i}$ and in $q_{L_i}\to q_{R_j}$ transitions there is a factor $m_{q_j}/v$. Notice that in $q_{R_i}\to q_{L_j}$ and in $q_{L_j}\to q_{R_i}$ transitions the factor is instead $m_{q_i}/v$. That is, in general, for a $q_j\to q_i$ vertex, there is a suppression factor given by the heaviest of the two quarks mass, $\max(m_{q_i},m_{q_j})$. This suppression factor is a very relevant one except, obviously, for any transition where the top quark is present.
\item A factor $\cab(\tti)$; from perturbative unitarity requirements on the scalar sector, is constrained to be at most one \cite{Akeroyd:2000wc,Ginzburg:2005dt,Botella:2015hoa,Grinstein:2015rtl,Cacchio:2016qyh}. Notice that in the limit $\cab\to 0$, the FCNC associated to the standard Higgs $\nh$ disappear, and the ones associated to the remaining scalars can be suppressed by making them heavier. Therefore, this is the quantity which, in a global approach, is bounded from above by Higgs mediated FCNC processes.
\item Finally, there is the factor $\un{{\rm [q]}i}\unC{{\rm [q]}j}$, which ranges from $0$ to $1/2$ (which is only reached when only \emph{one} transition is allowed). We remind that in BGL models, the analogous factor is $\V{iq}\Vc{jq}$. With the naive bounds on $\cab(\tti)$ from meson mixings in \cite{Botella:2015hoa}, one can compare and analyse how the suppression factors change in these new gBGL models. We collect the suppression factors for the different transitions in all BGL models in table \ref{TAB:FCNC:BGL} (with the corresponding power counting in Wolfenstein's parameter $\lambda$).
\begin{table}[h!]
\begin{center}
\begin{tabular}{|c||c|c|c|}\hline
\backslashbox{\footnotesize Transition}{\footnotesize Model} & $d$ & $s$ & $b$\\ \hline\hline
$u\leftrightarrow c$ & $\V{ud}\Vc{cd}\sim \lambda$ & $\V{us}\Vc{cs}\sim \lambda$ & $\V{ub}\Vc{cb}\sim \lambda^5$ \\ \hline
$u\leftrightarrow t$ & $\V{ud}\Vc{td}\sim \lambda^3$ & $\V{us}\Vc{ts}\sim \lambda^3$ & $\V{ub}\Vc{tb}\sim \lambda^3$ \\ \hline
$c\leftrightarrow t$ & $\V{cd}\Vc{td}\sim \lambda^4$ & $\V{cs}\Vc{ts}\sim \lambda^2$ & $\V{cb}\Vc{tb}\sim \lambda^2$ \\ \hline\hline
\backslashbox{\footnotesize Transition}{\footnotesize Model} & $u$ & $c$ & $t$\\ \hline\hline
$d\leftrightarrow s$ & $\Vc{ud}\V{us}\sim \lambda$ & $\Vc{cd}\V{cs}\sim \lambda$ & $\Vc{td}\V{ts}\sim \lambda^5$ \\ \hline
$d\leftrightarrow b$ & $\Vc{ud}\V{ub}\sim \lambda^3$ & $\Vc{cd}\V{cb}\sim \lambda^3$ & $\Vc{td}\V{tb}\sim \lambda^3$ \\ \hline
$s\leftrightarrow b$ & $\Vc{us}\V{ub}\sim \lambda^4$ & $\Vc{cs}\V{cb}\sim \lambda^2$ & $\Vc{ts}\V{tb}\sim \lambda^2$ \\ \hline
\end{tabular}
\caption{CKM suppression in FCNC transitions in BGL models.\label{TAB:FCNC:BGL}}
\end{center}
\end{table}
One has to compare the maximum value of $|\un{{\rm [q]}i}\unC{{\rm [q]}j}|_{\rm max}=1/2$ with $\lambda$, which corresponds to the less suppressed transition in some BGL models. This means that the most stringent constraint on BGL models obtained in \cite{Botella:2015hoa} for $|\cab(\tti)|$ should be reduced by a factor $2\lambda$. In this way, taking into account the analysis of \cite{Botella:2015hoa} and the constraints in \cite{Blankenburg:2012ex}, one can conclude that the gBGL models are safe, over the entire parameter space, provided
\begin{equation}
\ABS{\cab(\tti)}\leq 0.02\,.
\end{equation}
This constraint arises from $D^0$--$\bar D^0$ mixing and it turns out to be more stringent than the $K^0$--$\bar K^0$ one. It is worth mentioning that in some regions of the parameter space, this constraint will be relaxed. For example, in all BGL models this constraint is much weaker as shown in reference \cite{Botella:2015hoa}, and in some of them $|\cab(\tti)|$ can span the entire theoretically allowed parameter space, arriving to values of order $1$.
\end{itemize}
A final remark is related to the absence of the mass suppression factor on $t\leftrightarrow q$ transitions in gBGL models. In this case, the relevant bounds come from rare top decays $t\to\nh u,\nh c$, which give \cite{Botella:2015hoa}
\begin{equation}\label{eq:thq:00}
\text{BR}(t\to \nh q)=0.13\,\frac{\ABS{\un{{\rm [u]}q}\unC{{\rm [u]}t}}^2}{|\V{tb}|^2}\left(\cab(\tti)\right)^2\,.
\end{equation}
Considering the experimental bounds from ATLAS \cite{Aad:2014dya} and CMS \cite{Khachatryan:2014jya,Khachatryan:2016atv}, this yields
\begin{equation}
\ABS{\cab(\tti)}\leq 0.4\,,
\end{equation}
for the maximal value\footnote{The maximal value of $|\un{{\rm [u]}q}\unC{{\rm [u]}t}|$ cannot be obtained, obviously, for both $t\to \nh u$ and $t\to\nh c$ simultaneously.} $|\un{{\rm [u]}q}\unC{{\rm [u]}t}|=1/2$.\\
We conclude that the whole class of gBGL models does not lead to much larger flavour changing transitions than the ones arising in BGL models. This is in spite of having simultaneously FCNC in the up and in the down sectors. In the next sections we discuss other important differences.
%
\section{Near the top and the bottom models\label{SEC:neartb}}
BGL top and bottom models are the only renormalizable 2HDM that verify the MFV principle in any of the different versions one can find in the literature \cite{Botella:2009pq}. We devote this section to analyse in more detail the properties of gBGL models that are ``close'' to these models, that is, that they depart from the top or bottom models by a ``small amount''. Let us recall that the FCNC in the down and the up sectors of the top model are controlled by
\begin{equation}
(\mND)_{ij}=(\tti)\,\hat t^{\phantom{\ast}}_{{\rm [d]}i}\hat t^{\ast}_{{\rm [d]}j}\,m_{d_j}\,,\qquad
(\mNU)_{ij}=(\tti)\,\hat t^{\phantom{\ast}}_{{\rm [u]}i}\hat t^{\ast}_{{\rm [u]}j}\,m_{u_j}\,,
\end{equation}
and, following \refEQS{eq:uBGL:up}--\eqref{eq:uBGL:down}, $\hat t_{{\rm [u]}}$ and $\hat t_{{\rm [d]}}=\CKMd\hat t_{{\rm [u]}}$, are
\begin{equation}
\hat t_{{\rm [u]}}=\begin{pmatrix}0\\ 0\\ 1\end{pmatrix}\,,\quad
\hat t_{{\rm [d]}}=\begin{pmatrix}\Vc{td}\\ \Vc{ts}\\ \Vc{tb}\end{pmatrix}\,.
\end{equation}
It is clear that the BGL top model does not have FCNC in the up sector. Let us now consider small deviations from $\hat t_{\rm [d]}$ near the top model parameterised in terms of a complex vector $\vec \delta=(\delta_d, \delta_s,\delta_b)$ (with the appropriate normalisation):
\begin{equation}\label{eq:neartop:00}
\hat t_{\rm [d]}+\hat{\delta t}_{\rm [d]}\sim \begin{pmatrix}\Vc{td}(1+\delta_d)\\ \Vc{ts}(1+\delta_s)\\ \Vc{tb}(1+\delta_s)\end{pmatrix}\,.
\end{equation}
The elements of $\hat t_{\rm [d]}+\hat{\delta t}_{\rm [d]}$ control the flavour structure of the New Physics contributions to $K^0$--$\bar K^0$, $B^0_d$--$\bar B^0_d$ and $B^0_s$--$\bar B^0_s$. In particular, the leading order contributions to the different meson mixings $M_{12}$ have the following form:
\begin{align}
M_{12}[K^0]&\propto (\Vc{td}\V{ts})^2[1+2(\delta_s^\ast+\delta_d)]\,,\nonumber\\
M_{12}[B^0_d]&\propto (\Vc{td}\V{tb})^2[1+2(\delta_b^\ast+\delta_d)]\,,\nonumber\\
M_{12}[B^0_s]&\propto (\Vc{ts}\V{tb})^2[1+2(\delta_b^\ast+\delta_s)]\,.
\end{align}
Therefore, taking into account the phases of the dominant terms $(\Vc{ti}\V{tj})^2$, we can conclude that these ``near top'' models will give the same contribution to meson mixings provided
\begin{equation}
\re{\delta_d}\sim\re{\delta_s}\sim\im{\delta_s}\leq\mathcal O(\lambda^2),\ \text{and}\ \im{\delta_d}\sim\im{\delta_b}\leq\mathcal O(\lambda^3)\,.
\end{equation}
Notice that we are not stating that these models do not have any contribution to the meson mixings, the point is, rather, that these models are ``like the top BGL'' in the sense that they give the same contributions to $K^0$--$\bar K^0$, $B^0_d$--$\bar B^0_d$ and $B^0_s$--$\bar B^0_s$ mixings.
The immediate question is then: do these models produce too strong FCNC in the up sector, in particular in $D^0$--$\bar D^0$ or in top decays?
FCNC in the up sector are controlled by
\begin{equation}
\hat t_{\rm [u]}+\hat{\delta t}_{\rm [u]}=\CKMd(\hat t_{\rm [d]}+\hat{\delta t}_{\rm [d]})\sim \begin{pmatrix} \mathcal O(\lambda^5)\\ \delta_b\V{cb}\Vc{tb}\\ 1+\delta_b\end{pmatrix}\,.
\end{equation}
It is clear that $M_{12}[D^0]$ will have a suppression given by
\begin{equation}
(\delta_b\lambda^5\V{cb}\Vc{tb})^2\leq \lambda^{18}\,,
\end{equation}
much smaller than any of the contributions in dBGL models.\\
To analyse $t\to \nh c$ we have to compare the maximal value $1/4$ of $\ABS{\unu{q}\unuC{t}}^2$ in \refEQ{eq:thq:00} with the value obtained in the present case, $\ABS{(1+\delta_b)\delta_b\V{cb}\Vc{tb}}^2\sim\mathcal O(\lambda^8)$. The conclusion is evident in the whole parameter space: these models will produce $t\to \nh c$ still below the actual experimental bounds. The same conclusion applies to $t\to \nh u$. In the next section we will see that these models can depart in a sizable way from the top BGL model in different physical observables.\\ 
The bottom dBGL model is specified by $\hat b$ (\refEQS{eq:dBGL:down} and \eqref{eq:dBGL:up}):
\begin{equation}
\hat b_{\rm[d]}=\begin{pmatrix} 0\\ 0\\ 1\end{pmatrix}\,,\quad
\hat b_{\rm[u]}=\begin{pmatrix}\V{ub}\\ \V{cb}\\ \V{tb}\end{pmatrix}\,.
\end{equation}
This model generates FCNC in the up sector, and in particular it contributes to $D^0$--$\bar D^0$ mixing but not to $K^0$--$\bar K^0$, $B_d^0$--$\bar B_d^0$ and $B_s^0$--$\bar B_s^0$; gBGL models close to the $\hat b$ model that keep its essential properties are obtained with
\begin{equation}\label{eq:nearbottom:00}
\hat b_{\rm [u]}+\hat{\delta b}_{\rm [u]}\sim \begin{pmatrix}\V{ub}(1+\delta_u)\\ \V{cb}(1+\delta_c)\\ \V{tb}(1+\delta_t)\end{pmatrix}\,,
\end{equation}
where
\begin{equation}\label{eq:near:bottom:00}
\delta_u\sim\delta_c\sim\delta_t\leq \mathcal O(\lambda^2)\,,
\end{equation}
as before. To see what happens with the important constraints in the down sector, we show, as before, that $\hat b_{\rm [d]}+\hat{\delta b}_{\rm [d]}=\CKM(\hat b_{\rm [u]}+\hat{\delta b}_{\rm [u]})$,
\begin{equation}
\hat b_{\rm [d]}+\hat{\delta b}_{\rm [d]}\sim
\begin{pmatrix}\Vc{td}\V{tb}(\delta_t-\delta_c)+\Vc{ud}\V{ub}(\delta_u-\delta_c)\\ \Vc{ts}\V{tb}(\delta_t-\delta_c)\\ |\V{tb}|^2(1+\delta_t)\end{pmatrix}
\end{equation}
is the relevant quantity. With values as in \refEQ{eq:near:bottom:00}, it turns out that the contributions to the mixing in the down sector are much smaller than in any uBGL model. Nevertheless, we will also see, in the next section, that there are important differences with respect to the bottom dBGL model in other observables while considering the same kind of parameter values close to the bottom model.

\section{Baryon Asymmetry of the Universe\label{SEC:bau}}
\subsection{The leading gBGL contribution\label{sSEC:bau:gBGL}}

The presence of additional sources of flavour and CP violation in the gBGL models can enhance the contribution to the Baryonic Asymmetry of the Universe (BAU) with respect to SM expectations. Having $\mNU$ and $\mND$ in addition to $\mMU$ and $\mMD$, a weak basis invariant with a non-zero imaginary part already arises \cite{Botella:2012ab} at the 4$^{\text{th}}$ order\footnote{To be compared with the 12$^{\text{th}}$ order usual one, $-\frac{i}{6}\TR{[\mMU\mMUd,\mMD\mMDd]^3}$ \cite{Bernabeu:1986fc}.}\footnote{The rephasing invariance in the Higgs sector \cite{Botella:1994cs} imposes that the complete invariant should include the $H_2^\dagger H_1$ coefficient in the Lagrangian: $(\mu_{11}^2-\mu_{22}^2)\sb\cb$ \cite{Davidson:2005cw}. This does not introduce new phases.}:
\begin{equation}\label{eq:BAU:tr:00}
\iTR{\wND\wMDd\wMU\wMUd}\,.
\end{equation}
Considering $\wND$ in \refEQ{eq:N:ProjX3:00:1},
\begin{equation}\label{eq:BAU:tr:01}
\iTR{\wND\wMDd\wMU\wMUd}=(\tti)\iTR{\PRX{X 3}{q_L}\wMD\wMDd\wMU\wMUd}\,.
\end{equation}
For $\PRX{X 3}{q_L}=\PRX{\mathcal U^\prime 3}{d_L}$, that is, in the down parametrisation, \refEQ{eq:ProjX3:down:00},
\begin{multline}\label{eq:BAU:tr:02}
(\tti)\iTR{\UdL(\mathcal U^\prime\PR{3}\mathcal U^{\prime\dagger})\UdLd \UdL\mMD\mMDd\UdLd\UuL\mMU\mMUd\UuLd}=\\
(\tti)\iTR{(\mathcal U^\prime\PR{3}\mathcal U^{\prime\dagger})\mMD\mMDd\CKMd\mMU\mMUd\CKM}\,.
\end{multline}
Notice that, according to \refEQ{eq:MassBasis:01}, $\mMU\mMUd=\text{diag}(m_{u_j}^2)=\text{diag}(m_u^2,m_c^2,m_t^2)$, $\mMD\mMDd=\text{diag}(m_{d_j}^2)=\text{diag}(m_d^2,m_s^2,m_b^2)$. Following \refEQ{eq:n:vec:00},
\begin{equation}
(\mathcal U^\prime\PR{3}\mathcal U^{\prime\dagger})_{ij}=\und{i}\undC{j}\,,
\end{equation}
and thus
\begin{equation}\label{eq:BAU:tr:03}
\TR{(\mathcal U^\prime\PR{3}\mathcal U^{\prime\dagger})\mMD\mMDd\CKMd\mMU\mMUd\CKM}=\sum_{i,j,k}m_{d_j}^2m_{u_k}^2\und{i}\undC{j}\V{ki}\Vc{kj}\,,
\end{equation}
to obtain, finally
\begin{equation}\label{eq:BAU:tr:06}
\iTR{\wND\wMDd\wMU\wMUd}=\frac{i}{2}(\tti)\sum_{i,j,k}(m_{d_i}^2-m_{d_j}^2)m_{u_k}^2\und{i}\undC{j}\V{ki}\Vc{kj}\,.
\end{equation}
Although the discussion has relied on the use of the down parametrisation (from \refEQ{eq:BAU:tr:02} onwards), completely analogous results are obtained if the up parametrisation is used instead.\\
In order to estimate the enhancement in the BAU due to $\iTR{\wND\wMDd\wMU\wMUd}$, let us first retain only potentially leading contributions in terms of masses and powers of the Wolfenstein parameter $\lambda$ \cite{Wolfenstein:1983yz},
\begin{multline}\label{eq:BAU:estimate:00}
\iTR{\wND\wMDd\wMU\wMUd}\simeq \frac{i}{2}(\tti)\Big\{
\und{2}\undC{1}m_s^2\left[m_c^2\V{cs}\Vc{cd}+m_t^2\V{ts}\Vc{td}\right]\\
+\und{3}\undC{1}m_b^2\left[m_c^2\V{cb}\Vc{cd}+m_t^2\V{tb}\Vc{td}\right]
+\und{3}\undC{2}m_b^2\left[m_c^2\V{cb}\Vc{cs}+m_t^2\V{tb}\Vc{ts}\right]\Big\}\,
\end{multline}
that is,
\begin{multline}\label{eq:BAU:estimate:01}
\iTR{\wND\wMDd\wMU\wMUd}\sim i(\tti)\Big\{
\und{2}\undC{1}m_s^2\left[m_c^2\lambda+m_t^2\lambda^5\right]\\
+\und{3}\undC{1}m_b^2\left[m_c^2\lambda^3+m_t^2\lambda^3\right]
+\und{3}\undC{2}m_b^2\left[m_c^2\lambda^2+m_t^2\lambda^2\right]\Big\}\,.
\end{multline}
In the SM, the BAU is proportional to \cite{Trodden:1998ym,Morrissey:2012db}
\begin{equation}\label{eq:bau:SMvalue}
\text{BAU}_{\rm SM}\sim \frac{m_t^4 m_b^4 m_c^2 m_s^2}{E^{12}}\,J\,,
\end{equation}
where $J$ is the rephasing invariant imaginary part of the CKM quartets \cite{Jarlskog:1985ht}, $J=\im{\Vc{cs}\V{ts}\Vc{tb}\V{cb}}\simeq 3\times 10^{-5}$ and $E\sim 100$ GeV an energy of the order of the electroweak scale one. In \refEQ{eq:BAU:estimate:01}, we have contributions like the last one, giving
\begin{equation}
\text{BAU}_{\rm gBGL}\sim (\tti)\frac{m_t^2m_b^2}{E^4}\im{\und{3}\undC{2}\V{tb}\Vc{ts}}\,.
\end{equation}
This simple enhancement estimate with respect to \refEQ{eq:bau:SMvalue}, in terms of $\alpha=\arg(\und{3}\undC{2}\V{tb}\Vc{ts})$ and $\abs{\und{3}\undC{2}}$ (which has a maximal value of $1/2$), is:
\begin{multline}\label{eq:bau:enhancement}
\frac{\text{BAU}_{\rm gBGL}}{\text{BAU}_{\rm SM}}\sim (\tti) \ABS{\und{3}\undC{2}} \sin\alpha\,\frac{|\V{ts}|}{J}\frac{E^8}{m_t^2m_b^2m_c^2m_s^2}\\ 
\sim 10^{16} (\tti) \ABS{\und{3}\undC{2}} \sin\alpha\,,
\end{multline}
showing that there is margin for substantial enhancement of the BAU with respect to the SM, and with respect to BGL models too \cite{Botella:2012ab}. In appendix \ref{SEC:APP:edms} it is shown that we do not expect relevant constraints coming from electric dipole moments (EDM). 

\subsection{The vanishing BGL limits\label{sSEC:bau:BGL}}
In BGL models, the previous contribution is vanishing: let us explicitely check this known result \cite{Botella:2012ab}. For dBGL models, with $\un{\rm d}$ in \refEQ{eq:dBGL:down}, only one component is non-vanishing and $\iTR{\wND\wMDd\wMU\wMUd}=0$. The situation is slightly more involved for uBGL models. From \refEQ{eq:uBGL:down}, for an uBGL model of type $q$, $\und{i}=\Vc{qi}$. Then, going back to \refEQ{eq:BAU:tr:03},
\begin{equation}\label{eq:BAU:tr:07}
\TR{(\mathcal U^\prime\PR{3}\mathcal U^{\prime\dagger})\mMD\mMDd\CKMd\mMU\mMUd\CKM}=\sum_{i,j,k}m_{d_j}^2m_{u_k}^2\Vc{qi}\V{qj}\V{ki}\Vc{kj}\,,\quad (\text{uBGL } q)\,.
\end{equation}
Since, from unitarity of $\CKM$, $\sum_i\Vc{qi}\V{ki}=\delta_{qk}$, \refEQ{eq:BAU:tr:07} gives
\begin{equation}\label{eq:BAU:tr:08}
\TR{(\mathcal U^\prime\PR{3}\mathcal U^{\prime\dagger})\mMD\mMDd\CKMd\mMU\mMUd\CKM}=\sum_{j}m_{d_j}^2m_{u_q}^2|\V{qj}|^2\,,\quad (\text{uBGL } q)\,,
\end{equation}
and thus $\iTR{\wND\wMDd\wMU\wMUd}=0$.

\subsection{Rephasing invariance\label{sSEC:bau:rephasing}}
Although we have already mentioned that there are only two physical phases in $\un{\rm [d]}$ or $\un{\rm [u]}$, it is important to check that the invariant in \refEQ{eq:BAU:tr:06} is, at it should, invariant under individual rephasings of the different quark fields:
\begin{alignat}{3}
& d\mapsto e^{i\varphi_1}d,\quad & s\mapsto e^{i\varphi_2}s,\quad & b\mapsto e^{i\varphi_3}b,\nonumber\\
& \V{qd}\mapsto e^{i\varphi_1}\V{qd},\quad & \V{qs}\mapsto e^{i\varphi_2}\V{qs},\quad & \V{qb}\mapsto e^{i\varphi_3}\V{qb}\,.\label{eq:rephasing:00}
\end{alignat}
The origin of the rephasing of the CKM matrix is straightforward: since $\CKM=\UuLd\UdL$, the rephasing of the down quarks corresponds to
\begin{equation}
\UdL\mapsto \UdL\,\begin{pmatrix}e^{i\varphi_1} & 0 & 0\\ 0 & e^{i\varphi_2} & 0\\ 0 & 0 & e^{i\varphi_3}\end{pmatrix}\,.
\end{equation}
Then, since in \refEQ{eq:BAU:tr:02} $\PRX{\mathcal U^\prime 3}{d_L}=\UdL(\mathcal U^\prime\PR{3}\mathcal U^{\prime\dagger})\UdLd$,
\begin{equation}
\mathcal U^\prime\mapsto \begin{pmatrix}e^{-i\varphi_1} & 0 & 0\\ 0 & e^{-i\varphi_2} & 0\\ 0 & 0 & e^{-i\varphi_3}\end{pmatrix}\,\mathcal U^\prime
\end{equation}
to keep $\UdL\mathcal U^\prime$ invariant; as a consequence, under the rephasing in \refEQ{eq:rephasing:00},
\begin{equation}
\und{j}\mapsto e^{-i\varphi_j}\und{j}\,,
\end{equation}
and \refEQ{eq:BAU:tr:06} is clearly rephasing invariant.

\subsection{Enhancements in models near the top and the bottom models\label{sSEC:bau:neartb}}
The BAU generated in the top BGL model is proportional to \cite{Botella:2012ab}
\begin{multline}
\iTR{\mMD\mNDd\mMD\mMDd\mMU\mMUd\mMD\mMDd}\\ =-(\tti)(m_b^2-m_s^2)(m_b^2-m_d^2)(m_s^2-m_d^2)(m_c^2-m_u^2)\im{\Vc{cs}\V{ts}\Vc{tb}\V{cb}}\,,
\end{multline}
in such a way that the ratio to the SM BAU is, for $E\sim 100$ GeV,
\begin{equation}
\frac{\text{BAU}_{\rm BGL-t}}{\text{BAU}_{\rm SM}}=(\tti)\frac{E^4}{m_t^4}\sim 1\,.
\end{equation}
Therefore, as far as CP violation is concerned, the top model suffers the same problem as the SM in not being able to generate sufficient BAU. There is no enhancement in the top BGL 2HDM. We can now consider a small departure from the top BGL model as in \refEQ{eq:neartop:00} and apply it to \refEQ{eq:bau:enhancement}. With
\begin{equation}
\undC{3}\sim\Vc{tb}(1+\delta_b)\,,\quad \undC{2}\sim\V{ts}(1+\delta_s^\ast)\,,
\end{equation}
\begin{equation}
\im{\undC{2}\undC{3}\V{tb}\Vc{ts}}\sim |\V{ts}|^2\im{\delta_b+\delta_s^\ast}\,,
\end{equation}
and \refEQ{eq:neartop:00} gives
\begin{equation}
\frac{\text{BAU}_{\rm near\ t}}{\text{BAU}_{\rm SM}}=10^{16}(\tti)|\V{ts}|\im{\delta_b+\delta_s^\ast}\,,
\end{equation}
that can produce an enhancement as large as $10^{12}$. Even if we are close to a top BGL model we find that, contrary to what happens with top model itself, there can be enough CP violation in this class of models to generate the BAU.
In an analogous way, the BAU generated in the bottom BGL model is proportional to
\begin{multline}
\iTR{\mMU\mNUd\mMU\mMUd\mMD\mMDd\mMU\mMUd}\\ =-(\tti)(m_t^2-m_c^2)(m_t^2-m_u^2)(m_c^2-m_u^2)(m_s^2-m_d^2)\im{\Vc{cs}\V{ts}\Vc{tb}\V{cb}}\,,
\end{multline}
and the ratio to the SM BAU, for $E\sim 100$ GeV, is
\begin{equation}
\frac{\text{BAU}_{\rm BGL-b}}{\text{BAU}_{\rm SM}}=(\tti)\frac{E^4}{m_b^4}\sim 10^5\,.
\end{equation}
In a pure bottom model there is still not enough CP violation -- coming from the Yukawa sector -- to generate the BAU. But, if we depart from an exact bottom BGL model in the manner explained in \refEQ{eq:nearbottom:00}, with
\begin{equation}
\und{3}\sim|\V{tb}|^2(1+\delta_t)\,,\quad \undC{2}\sim\V{ts}\Vc{tb}(\delta_t^\ast-\delta_c^\ast)\,,
\end{equation}
then
\begin{equation}
\im{\und{3}\undC{2}\V{tb}\Vc{ts}}\sim |\V{tb}|^4|\V{ts}|^2\im{\delta_t^\ast-\delta_c^\ast}\,,
\end{equation}
and we obtain, for the models near the bottom BGL one, 
\begin{equation}
\frac{\text{BAU}_{\rm near\ b}}{\text{BAU}_{\rm SM}}=10^{16}(\tti)|\V{ts}|\im{\delta_t^\ast-\delta_c^\ast}\,,
\end{equation}
that is, a potential enhancement of $10^{13}$ with respect to the SM.


\clearpage
\section*{Conclusions\label{SEC:conclusions}}
We have analysed the question of FCNC in the framework of 2HDM. In this paper, it has been shown that BGL models can be readily generalized to gBGL models which keep the nice feature of renormalizability, but no longer belong to the Minimal Flavour Violation Framework.
Contrary to BGL models, gBGL models contain four extra flavour parameters beyond the CKM matrix $\CKM$, in spite of the drastic reduction in the number of parameters in gBGL, when compared to those present in the general 2HDM. 
It has been shown that in gBGL models FCNC are present at tree level, but in a controlled manner, rendering them plausible extensions of the SM. In fact they can have simultaneously FCNC in the up and in the down sector, a difference with BGL models that have FCNC in one of the sectors only.

The flavour structure of Yukawa couplings in gBGL is achieved, in a natural way, through the introduction of a $\ZZ$ symmetry, at the Lagrangian level. So gBGL models use the same symmetry as proposed by Glashow and Weinberg in NFC models, the only difference lies in the way the quark fields transform under $\ZZ$. gBGL models contain BGL models as special cases, in the sense that in the parameter space of gBGL there are regions where one comes close to particular BGL models. Note that in the limit where a BGL is reached, the Lagrangian acquires a larger symmetry, namely a $\mathbb{Z}_4$ or a $U(1)$ symmetry, dictated by the corresponding neutrino type (either Majorana or Dirac) of the BGL model.

It is well known that the SM does not generate sufficient Baryon Asymmetry of the Universe (BAU) at the electroweak phase transition. One of the reasons for this is the fact that in the SM CP violation is too small. This stems from the fact that the WB invariant \cite{Jarlskog:1985ht,Bernabeu:1986fc} controlling CP violation in the SM is of order mass to the 12$^{\text{th}}$. We have analysed in detail how lower order CP odd invariants appear in gBGL models. It turns out that they arise at a much lower mass order which leads to the possibility of generating a much higher BAU, compared to that in the SM.

\section*{Acknowledgments}
This work is partially supported by Spanish MINECO under grant FPA2015-68318-R and by the Severo Ochoa Excellence Center Project SEV-2014-0398, by Generalitat Valenciana under grant GVPROMETEOII 2014-049 and by Funda\c{c}\~ao para a Ci\^encia e a Tecnologia (FCT, Portugal) through the projects CERN/FIS-NUC/0010/2015 and CFTP-FCT Unit 777 (UID/FIS/00777/2013) which are partially funded through POCTI (FEDER), COMPETE, QREN and EU. MN acknowledges support from FCT through postdoctoral grant SFRH/BPD/112999/2015.


\clearpage
\appendix

\section{Necessary and Sufficient Conditions for gBGL\label{SEC:APP:cond}}
We complete in this appendix the proof of the sufficient conditions in the following general result: \emph{the WB invariant matrix conditions}
\begin{alignat}{3}
&\Ydd{2}\Yd{1}=0,&\quad & \Ydd{2}\Yu{1}=0,\nonumber\\
&\Yud{2}\Yu{1}=0,&\quad & \Yud{2}\Yd{1}=0,\tag{\ref{eq:gBGLcond:00}}
\end{alignat}
\emph{are the necessary and sufficient conditions to define gBGL models or a type I 2HDM\footnote{Since we are not specifying the leptonic sector, with type I we also refer to type X 2HDM.}, provided there are no massless quarks.}\\ 
It is always possible, in general, to write
\begin{equation}\label{eq:diagdecomp:00}
\Yd{i}=W_{d_i}\,D_{d_i}\,U_{d_i}^\dagger,\qquad \Yu{i}=W_{u_i}\,D_{u_i}\,U_{u_i}^\dagger,
\end{equation}
where $W_{d_i}$, $W_{u_i}$, $U_{d_i}$ and $U_{u_i}$ are unitary matrices, and $D_{d_i}$ and $D_{u_i}$ are diagonal ones.
From $\Yud{2}\Yu{1}=0$ it is straightforward that $[\Yu{1}\Yud{1},\Yu{2}\Yud{2}]=0$ and thus one can choose
\begin{equation}
W_{u_1}=W_{u_2}=W_{u}\,,
\end{equation}
while from $\Ydd{2}\Yd{1}=0$ it follows that $[\Yd{1}\Ydd{1},\Yd{2}\Ydd{2}]=0$ and therefore we can also choose
\begin{equation}
W_{d_1}=W_{d_2}=W_{d}\,.
\end{equation}
Now, in $\Yud{2}\Yu{1}=0$, $W_{u}$ simplifies away and we have
\begin{equation}\label{eq:cond:diagup:00}
D_{u_2}^\dagger\,D_{u_1}=0\,,
\end{equation}
and similarly, for $\Ydd{2}\Yd{1}=0$,
\begin{equation}\label{eq:cond:diagdown:00}
D_{d_2}^\dagger\,D_{d_1}=0\,.
\end{equation}
If there are no massless quarks, there are two kinds of solutions for \refEQ{eq:cond:diagup:00}, 
\begin{itemize}
\item[(a)] type I 2HDM
\begin{equation}\label{eq:typeIup:00}
D_{u_1}=\begin{pmatrix}0&0&0\\ 0&0&0\\ 0&0&0\end{pmatrix},\quad D_{u_2}=\begin{pmatrix}u_1&0&0\\ 0&u_2&0\\ 0&0&u_3\end{pmatrix}\,,
\end{equation}
with $u_i\neq 0$ in order to have massive up quarks. Notice that interchanging $D_{u_1}\leftrightarrows D_{u_2}$ will give rise to the same model, as explained later.
\item[(b)] gBGL
\begin{equation}\label{eq:gBGLup:00}
D_{u_1}=\begin{pmatrix}u_1&0&0\\ 0&0&0\\ 0&0&0\end{pmatrix},\quad D_{u_2}=\begin{pmatrix}0&0&0\\ 0&u_2&0\\ 0&0&u_3\end{pmatrix}\,,
\end{equation}
with $u_i\neq 0$ again. As above, exchanging $D_{u_1}\leftrightarrows D_{u_2}$ does not introduce new models. We should also take into account the possibility that $u_1\neq 0$ is in a different position in the diagonal of $D_{u_1}$ while respecting $D_{u_2}^\dagger D_{u_1}=0$, which is ensured with a corresponding permutation of the diagonal elements of $D_{u_2}$.
\end{itemize}
Similarly to the up sector, we have different solutions of \refEQ{eq:cond:diagdown:00}, and we consider three possibilities.
\begin{itemize}
\item[(a)] First,
\begin{equation}\label{eq:typeIdown:00}
D_{d_1}=\begin{pmatrix}0&0&0\\ 0&0&0\\ 0&0&0\end{pmatrix},\quad D_{d_2}=\begin{pmatrix}d_1&0&0\\ 0&d_2&0\\ 0&0&d_3\end{pmatrix}\,,
\end{equation}
with $d_i\neq 0$. Notice that \refEQ{eq:typeIup:00} together with \refEQ{eq:typeIdown:00} with interchanged $D_{d_1}\leftrightarrows D_{d_2}$ do not match in order to be a solution of \refEQS{eq:gBGLcond:00}.
\item[(b)] Second, $D_{d_1}$ and $D_{u_1}$ have equal rank, and we could consider in general permutations of the diagonal elements, for example
\begin{equation}\label{eq:gBGLdown:00}
D_{d_1}=\begin{pmatrix}0&0&0\\ 0&d_2&0\\ 0&0&0\end{pmatrix},\quad D_{d_2}=\begin{pmatrix}d_1&0&0\\ 0&0&0\\ 0&0&d_3\end{pmatrix}\,.
\end{equation}
Then, of course, the rank of $D_{d_2}$ is equal to the rank of $D_{u_2}$.
\item[(c)] Third, $D_{d_1}$ and $D_{u_1}$ have different rank (and therefore $D_{d_2}$ and $D_{u_2}$ also have different rank), for example
\begin{equation}\label{eq:gBGLdown:01}
D_{d_1}=\begin{pmatrix}d_1&0&0\\ 0&d_2&0\\ 0&0&0\end{pmatrix},\quad D_{d_2}=\begin{pmatrix}0&0&0\\ 0&0&0\\ 0&0&d_3\end{pmatrix}\,.
\end{equation}
\end{itemize}
We have to explore now which solutions to \refEQS{eq:gBGLcond:00} arise from the available possibilities in \refEQS{eq:typeIup:00}--\eqref{eq:gBGLup:00} and \refEQS{eq:typeIdown:00}--\eqref{eq:gBGLdown:01}.
\begin{itemize}
\item $D_{d_1}=0$ if and only if $D_{u_1}=0$, which corresponds to a type I or X 2HDM. The proof is simple: if $D_{d_1}=0$, $D_{d_2}$ has rank 3, and thus $\Yd{2}$ has rank 3. Since $\Ydd{2}\Yu{1}=0$, all column vectors of $\Yu{1}$ are in the null-space of $\Ydd{2}$ (they are all non-zero vectors transformed into the zero or null vector), but since $\text{rank}(\Ydd{2})=3$, according to the rank-nullity theorem, the null-space of $\Ydd{2}$ has dimension $3-3=0$, and thus $\Yu{1}=0$, that is $D_{u_1}=0$. \REFEQS{eq:gBGLcond:00} are then trivially verified. Of course, there is also the solution $D_{d_2}=D_{u_2}=0$, which is completely equivalent with a trivial relabelling of the scalar doublets $\Hd{1}\leftrightarrows\Hd{2}$.
\item Next we show that concerning \refEQ{eq:gBGLup:00} and \refEQS{eq:gBGLdown:00}--\eqref{eq:gBGLdown:01}, the ranks of the Yukawa matrices should match in the following manner: $\text{rank}(D_{d_1})=\text{rank}(D_{u_1})$ and $\text{rank}(D_{d_2})=\text{rank}(D_{u_2})$. Consider for definiteness $D_{u_1}$ and $D_{u_2}$ as in \refEQ{eq:gBGLup:00}. First, since $\Yud{2}\Yd{1}=0$ and $\Ydd{2}\Yu{1}=0$, with $W=W_u^\dagger W_d$,
\begin{equation}
D_{u_2}^\dagger\, W \, D_{d_1}=0,\qquad D_{d_2}^\dagger\, W^\dagger \, D_{u_1}=0\,.
\end{equation}
$D_{u_2}^\dagger W$ has rank 2, and thus its null-space has dimension 1; according to the first equation above, if $D_{d_1}$ had rank 2, then the null-space of $D_{u_2}^\dagger W$ would have at least dimension 2 in contradiction with the first statement: consequently, the rank of $D_{d_1}$ has to be 1. Then $\text{rank}(D_{d_1})=\text{rank}(D_{u_1})=1$ and  $\text{rank}(D_{d_2})=\text{rank}(D_{u_2})=2$. That is, the matrices in \refEQ{eq:gBGLdown:00} match the ones in \refEQ{eq:gBGLup:00} for solutions of \refEQS{eq:gBGLcond:00} while \refEQS{eq:gBGLdown:00} and \eqref{eq:gBGLup:00} do not match. Finally, relabelling of scalar doublets $\Hd{1}\leftrightarrows\Hd{2}$ gives equivalent solutions with $\text{rank}(D_{d_1})=\text{rank}(D_{u_1})=2$ and $\text{rank}(D_{d_2})=\text{rank}(D_{u_2})=1$.
\item The following step is to show that the most general case can be taken to be
\begin{alignat}{3}
&D_{u_1}=\begin{pmatrix}u_1&0&0\\ 0&0&0\\ 0&0&0\end{pmatrix},&\quad &D_{u_2}=\begin{pmatrix}0&0&0\\ 0&u_2&0\\ 0&0&u_3\end{pmatrix}\,,\nonumber\\
&D_{d_1}=\begin{pmatrix}d_1&0&0\\ 0&0&0\\ 0&0&0\end{pmatrix},&\quad &D_{d_2}=\begin{pmatrix}0&0&0\\ 0&d_2&0\\ 0&0&d_3\end{pmatrix}\,.\label{eq:gBGLdiag:00}
\end{alignat}
It is trivial to check that with the insertion of
\begin{equation}
P_{12}=\begin{pmatrix}0&1&0\\ 1&0&0\\ 0&0&1\end{pmatrix}\,,
\end{equation}
in $\Yd{i}=W_d\,P_{12}P_{12}\,D_{d_i}\,P_{12}P_{12}\,U_d^\dagger=W_d^\prime\,(P_{12}\,D_{d_i}\,P_{12})\,U_d^{\prime\dagger}$, one can redefine $W_d^\prime=W_d\,P_{12}$ and $U_d^{\prime}=U_d P_{12}$, while $P_{12}$ in $P_{12}\,D_{d_i}\,P_{12}$ permutes the first and second elements in the diagonal. This would bring, for example, \refEQ{eq:gBGLdown:00} to the desired form, matching with \refEQ{eq:gBGLup:00}. 
\item Finally, substituting \refEQ{eq:gBGLdiag:00} in $D_{u_1}^\dagger\, W \, D_{d_2}=0$,
\begin{equation}
u_1^\ast\,\begin{pmatrix}0& d_2\,W_{12} & d_3\,W_{13}\\ 0&0&0\\ 0&0&0\end{pmatrix}=0\,,
\end{equation}
we get $W_{12}=W_{13}=0$, while from $D_{u_2}^\dagger\, W \, D_{d_1}=0$,
\begin{equation}
d_1\,\begin{pmatrix} 0&0&0\\ u_2^\ast W_{21} &0&0\\ u_3^\ast W_{31} &0&0\end{pmatrix}=0\,,
\end{equation}
we have $W_{21}=W_{31}=0$. Then $W$ has the block structure
\begin{equation}
W=W_u^\dagger W_d=e^{i\omega}\begin{pmatrix}1&0&0\\ 0&W_{22}&W_{23}\\ 0&W_{32}&W_{33}\end{pmatrix}\,.
\end{equation}
One can now complete the proof; going back to \refEQ{eq:diagdecomp:00}:
\begin{align}
&\Yu{1}=W_u\begin{pmatrix}u_1&0&0\\ 0&0&0\\ 0&0&0\end{pmatrix}U_{u_1}^\dagger= W_u\begin{pmatrix}\times&\times&\times\\ 0&0&0\\ 0&0&0\end{pmatrix}\,,\nonumber\\
&\Yu{2}=W_u\begin{pmatrix}0&0&0\\ 0&u_2&0\\ 0&0&u_3\end{pmatrix}U_{u_2}^\dagger= W_u\begin{pmatrix}0&0&0\\ \times&\times&\times\\ \times&\times&\times\end{pmatrix}\,,
\end{align}
\begin{align}
&\Yd{1}=W_d\begin{pmatrix}d_1&0&0\\ 0&0&0\\ 0&0&0\end{pmatrix}U_{d_1}^\dagger= W_u\,W\begin{pmatrix}\times&\times&\times\\ 0&0&0\\ 0&0&0\end{pmatrix}= W_u\begin{pmatrix}\times&\times&\times\\ 0&0&0\\ 0&0&0\end{pmatrix}\,,\nonumber\\
&\Yd{2}=W_d\begin{pmatrix}0&0&0\\ 0&d_2&0\\ 0&0&d_3\end{pmatrix}U_{d_2}^\dagger= W_u\,W\begin{pmatrix}0&0&0\\ \times&\times&\times\\ \times&\times&\times\end{pmatrix}= W_u\,\begin{pmatrix}0&0&0\\ \times&\times&\times\\ \times&\times&\times\end{pmatrix}\,.
\end{align}
With a WB transformation given by $Q_L\mapsto W_u\,Q_L$ we arrive to the equivalent gBGL structures in \refEQS{eq:Yukawa:Texture:gBGL:00}.
\end{itemize}

\section{Electric Dipole Moments\label{SEC:APP:edms}}
Attending to the explicit form of the $\mND$ and $\mNU$ matrices in section \ref{SEC:paramet} and the discussion in section \ref{SEC:bau}, it is clear that gBGL models include new sources of CP violation. Although the dedicated analysis of section \ref{SEC:fcnc} addresses the controlled nature of the FCNC, one might still be concerned with the possibility that too large contributions to EDMs are present, induced in particular by the couplings to the charged scalar $\cHp$ or by the (flavour) diagonal couplings to the neutral scalars, present in \refEQS{eq:YukawasFull:00}.

For the latter, let us consider for the moment a generic generation $j$ of the up sector (for the down sector the reasoning is identical); the flavour diagonal couplings in \refEQS{eq:YukawasFull:00} are controlled by $[\mMU]_{jj}$ and $[\mNU]_{jj}\gR+[\mNUd]_{jj}\gL$ for the neutral scalars $\nHH$ and $\nHR$, and by $i[\mNU]_{jj}\gR-i[\mNUd]_{jj}\gL$ for the pseudoscalar $\nA$. Following \refEQ{eq:Nq:gen:00:2},
\begin{equation}
[\mNU]_{jj}=\tb m_{u_j}-(\tti)\unu{j}\unuC{j}m_{u_j}=(\tb-(\tti)|\unu{j}|^2) m_{u_j}=[\mNUd]_{jj}\, ,\label{eq:Nu:diag:00}
\end{equation}
and hence, for the coupling to $\nHR$,
\begin{equation}
[\mNU]_{jj}\gR+[\mNUd]_{jj}\gL=(\tb-(\tti)|\unu{j}|^2) m_{u_j}\,,
\end{equation}
 while for the coupling to $\nA$, 
\begin{equation}
i[\mNU]_{jj}\gR-i[\mNUd]_{jj}\gL=i(\tb-(\tti)|\unu{j}|^2) m_{u_j}\gamma_5\,.
\end{equation}
In general we have 
\begin{equation}
\im{[\mNU]_{jj}}=\im{[\mND]_{jj}}=0
\end{equation}
and
\begin{equation}
\iTR{\mNU\mMUd}=\iTR{\mND\mMDd}=0\,.
\end{equation}
With such couplings, the flavour diagonal quark interactions with the different neutral scalars are not CP violating \cite{Branco:1999fs,Nebot:2015wsa}, and no contribution to the EDMs arises. This applies to one loop contributions, to two loop Barr-Zee contributions \cite{Barr:1990vd}, to two loop contributions to the three gluon Weinberg operator \cite{Weinberg:1989dx} and to effective four fermion operators \cite{Jung:2013hka}. Notice that, although $\nHH$ and $\nHR$ are not the physical scalars, this conclusion remains unchanged when they are rotated into the mass eigenstates $\nH$ and $\nh$ in \refEQ{eq:NeutralScalars:02}.\\ 
For the charged scalar $\cHp$, one loop diagrams and contributions to the three gluon operator are not CP violating. Furthermore, the charged Higgs Barr-Zee contributions which are expected to be dominant \cite{BowserChao:1997bb} (see also \cite{Ilisie:2015tra}), are also CP conserving because their proportionality to $\im{[\mNU]_{tt}[\mNQ{q}]_{jj}}=0$.

\clearpage

\begin{thebibliography}{10}

\bibitem{Aad:2012tfa}
{\bf ATLAS} Collaboration, G.~Aad {\em et~al.}, {\it {Observation of a new
  particle in the search for the Standard Model Higgs boson with the ATLAS
  detector at the LHC}},  {\em Phys. Lett.} {\bf B716} (2012) 1--29,
  [\href{http://xxx.lanl.gov/abs/1207.7214}{{\tt 1207.7214}}].

\bibitem{Chatrchyan:2012xdj}
{\bf CMS} Collaboration, S.~Chatrchyan {\em et~al.}, {\it {Observation of a new
  boson at a mass of 125 GeV with the CMS experiment at the LHC}},  {\em Phys.
  Lett.} {\bf B716} (2012) 30--61,
  [\href{http://xxx.lanl.gov/abs/1207.7235}{{\tt 1207.7235}}].

\bibitem{Lee:1973iz}
T.~Lee, {\it {A Theory of Spontaneous T Violation}},  {\em Phys.Rev.} {\bf D8}
  (1973) 1226--1239.

\bibitem{Branco:2011iw}
G.~Branco, P.~Ferreira, L.~Lavoura, M.~Rebelo, M.~Sher, {\em et~al.}, {\it
  {Theory and phenomenology of two-Higgs-doublet models}},  {\em Phys.Rept.}
  {\bf 516} (2012) 1--102, [\href{http://xxx.lanl.gov/abs/1106.0034}{{\tt
  1106.0034}}].

\bibitem{Ivanov:2017dad}
I.~P. Ivanov, {\it {Building and testing models with extended Higgs sectors}},
  \href{http://xxx.lanl.gov/abs/1702.03776}{{\tt 1702.03776}}.

\bibitem{Glashow:1976nt}
S.~L. Glashow and S.~Weinberg, {\it {Natural Conservation Laws for Neutral
  Currents}},  {\em Phys.Rev.} {\bf D15} (1977) 1958.

\bibitem{Paschos:1976ay}
E.~Paschos, {\it {Diagonal Neutral Currents}},  {\em Phys.Rev.} {\bf D15}
  (1977) 1966.

\bibitem{Pich:2009sp}
A.~Pich and P.~Tuzon, {\it {Yukawa Alignment in the Two-Higgs-Doublet Model}},
  {\em Phys.Rev.} {\bf D80} (2009) 091702,
  [\href{http://xxx.lanl.gov/abs/0908.1554}{{\tt 0908.1554}}].

\bibitem{Mahmoudi:2009zx}
F.~Mahmoudi and O.~Stal, {\it {Flavor constraints on the two-Higgs-doublet
  model with general Yukawa couplings}},  {\em Phys. Rev.} {\bf D81} (2010)
  035016, [\href{http://xxx.lanl.gov/abs/0907.1791}{{\tt 0907.1791}}].

\bibitem{Crivellin:2013wna}
A.~Crivellin, A.~Kokulu, and C.~Greub, {\it {Flavor-phenomenology of
  two-Higgs-doublet models with generic Yukawa structure}},  {\em Phys.Rev.}
  {\bf D87} (2013), no.~9 094031,
  [\href{http://xxx.lanl.gov/abs/1303.5877}{{\tt 1303.5877}}].

\bibitem{Gaitan:2015hga}
R.~Gaitán, R.~Martinez, and J.~H.~M. de~Oca, {\it {Rare top decay $t
  \rightarrow c \gamma$ with flavor changing neutral scalar interactions in two
  Higgs doublet model}},  {\em Phys. Rev.} {\bf D94} (2016), no.~9 094038,
  [\href{http://xxx.lanl.gov/abs/1503.04391}{{\tt 1503.04391}}].

\bibitem{Altunkaynak:2015twa}
B.~Altunkaynak, W.-S. Hou, C.~Kao, M.~Kohda, and B.~McCoy, {\it {Flavor
  Changing Heavy Higgs Interactions at the LHC}},  {\em Phys. Lett.} {\bf B751}
  (2015) 135--142, [\href{http://xxx.lanl.gov/abs/1506.00651}{{\tt
  1506.00651}}].

\bibitem{Arhrib:2015maa}
A.~Arhrib, R.~Benbrik, C.-H. Chen, M.~Gomez-Bock, and S.~Semlali, {\it
  {Two-Higgs-doublet type-II and -III models and $t\rightarrow c h$ at the
  LHC}},  {\em Eur. Phys. J.} {\bf C76} (2016), no.~6 328,
  [\href{http://xxx.lanl.gov/abs/1508.06490}{{\tt 1508.06490}}].

\bibitem{Kim:2015zla}
C.~S. Kim, Y.~W. Yoon, and X.-B. Yuan, {\it {Exploring top quark FCNC within
  2HDM type III in association with flavor physics}},  {\em JHEP} {\bf 12}
  (2015) 038, [\href{http://xxx.lanl.gov/abs/1509.00491}{{\tt 1509.00491}}].

\bibitem{Enomoto:2015wbn}
T.~Enomoto and R.~Watanabe, {\it {Flavor constraints on the Two Higgs Doublet
  Models of Z$_{2}$ symmetric and aligned types}},  {\em JHEP} {\bf 05} (2016)
  002, [\href{http://xxx.lanl.gov/abs/1511.05066}{{\tt 1511.05066}}].

\bibitem{Benbrik:2015evd}
R.~Benbrik, C.-H. Chen, and T.~Nomura, {\it {$h,Z\to \ell_i \bar\ell_j$,
  $\Delta a_{\mu}$, $\tau\to (3\mu,\mu \gamma)$ in generic two-Higgs-doublet
  models}},  {\em Phys. Rev.} {\bf D93} (2016), no.~9 095004,
  [\href{http://xxx.lanl.gov/abs/1511.08544}{{\tt 1511.08544}}].

\bibitem{Branco:1996bq}
G.~Branco, W.~Grimus, and L.~Lavoura, {\it {Relating the scalar flavor changing
  neutral couplings to the CKM matrix}},  {\em Phys.Lett.} {\bf B380} (1996)
  119--126, [\href{http://xxx.lanl.gov/abs/hep-ph/9601383}{{\tt
  hep-ph/9601383}}].

\bibitem{Buras:2000dm}
A.~Buras, P.~Gambino, M.~Gorbahn, S.~Jager, and L.~Silvestrini, {\it {Universal
  unitarity triangle and physics beyond the standard model}},  {\em Phys.Lett.}
  {\bf B500} (2001) 161--167,
  [\href{http://xxx.lanl.gov/abs/hep-ph/0007085}{{\tt hep-ph/0007085}}].

\bibitem{D'Ambrosio:2002ex}
G.~D'Ambrosio, G.~Giudice, G.~Isidori, and A.~Strumia, {\it {Minimal flavor
  violation: An Effective field theory approach}},  {\em Nucl.Phys.} {\bf B645}
  (2002) 155--187, [\href{http://xxx.lanl.gov/abs/hep-ph/0207036}{{\tt
  hep-ph/0207036}}].

\bibitem{Cirigliano:2005ck}
V.~Cirigliano, B.~Grinstein, G.~Isidori, and M.~B. Wise, {\it {Minimal flavor
  violation in the lepton sector}},  {\em Nucl.Phys.} {\bf B728} (2005)
  121--134, [\href{http://xxx.lanl.gov/abs/hep-ph/0507001}{{\tt
  hep-ph/0507001}}].

\bibitem{Botella:2009pq}
F.~Botella, G.~Branco, and M.~Rebelo, {\it {Minimal Flavour Violation and
  Multi-Higgs Models}},  {\em Phys.Lett.} {\bf B687} (2010) 194--200,
  [\href{http://xxx.lanl.gov/abs/0911.1753}{{\tt 0911.1753}}].

\bibitem{Lavoura:1994ty}
L.~Lavoura, {\it {Models of CP violation exclusively via neutral scalar
  exchange}},  {\em Int.J.Mod.Phys.} {\bf A9} (1994) 1873--1888.

\bibitem{Botella:2011ne}
F.~Botella, G.~Branco, M.~Nebot, and M.~Rebelo, {\it {Two-Higgs Leptonic
  Minimal Flavour Violation}},  {\em JHEP} {\bf 1110} (2011) 037,
  [\href{http://xxx.lanl.gov/abs/1102.0520}{{\tt 1102.0520}}].

\bibitem{Botella:2014ska}
F.~Botella, G.~Branco, A.~Carmona, M.~Nebot, L.~Pedro, and M.~Rebelo, {\it
  {Physical Constraints on a Class of Two-Higgs Doublet Models with FCNC at
  tree level}},  {\em JHEP} {\bf 1407} (2014) 078,
  [\href{http://xxx.lanl.gov/abs/1401.6147}{{\tt 1401.6147}}].

\bibitem{Bhattacharyya:2014nja}
G.~Bhattacharyya, D.~Das, and A.~Kundu, {\it {Feasibility of light scalars in a
  class of two-Higgs-doublet models and their decay signatures}},  {\em
  Phys.Rev.} {\bf D89} (2014) 095029,
  [\href{http://xxx.lanl.gov/abs/1402.0364}{{\tt 1402.0364}}].

\bibitem{Botella:2015hoa}
F.~J. Botella, G.~C. Branco, M.~Nebot, and M.~N. Rebelo, {\it {Flavour Changing
  Higgs Couplings in a Class of Two Higgs Doublet Models}},  {\em Eur. Phys.
  J.} {\bf C76} (2016), no.~3 161,
  [\href{http://xxx.lanl.gov/abs/1508.05101}{{\tt 1508.05101}}].

\bibitem{Georgi:1978ri}
H.~Georgi and D.~V. Nanopoulos, {\it {Suppression of Flavor Changing Effects
  From Neutral Spinless Meson Exchange in Gauge Theories}},  {\em Phys. Lett.}
  {\bf B82} (1979) 95.

\bibitem{Donoghue:1978cj}
J.~F. Donoghue and L.~F. Li, {\it {Properties of Charged Higgs Bosons}},  {\em
  Phys. Rev.} {\bf D19} (1979) 945.

\bibitem{Botella:1994cs}
F.~J. Botella and J.~P. Silva, {\it {Jarlskog - like invariants for theories
  with scalars and fermions}},  {\em Phys. Rev.} {\bf D51} (1995) 3870--3875,
  [\href{http://xxx.lanl.gov/abs/hep-ph/9411288}{{\tt hep-ph/9411288}}].

\bibitem{Ferreira:2010ir}
P.~M. Ferreira and J.~P. Silva, {\it {Abelian symmetries in the
  two-Higgs-doublet model with fermions}},  {\em Phys. Rev.} {\bf D83} (2011)
  065026, [\href{http://xxx.lanl.gov/abs/1012.2874}{{\tt 1012.2874}}].

\bibitem{Branco:1985aq}
G.~C. Branco and M.~N. Rebelo, {\it {The Higgs Mass in a Model With Two Scalar
  Doublets and Spontaneous {CP} Violation}},  {\em Phys. Lett.} {\bf B160}
  (1985) 117--120.

\bibitem{Akeroyd:2000wc}
A.~G. Akeroyd, A.~Arhrib, and E.-M. Naimi, {\it {Note on tree level unitarity
  in the general two Higgs doublet model}},  {\em Phys. Lett.} {\bf B490}
  (2000) 119--124, [\href{http://xxx.lanl.gov/abs/hep-ph/0006035}{{\tt
  hep-ph/0006035}}].

\bibitem{Ginzburg:2005dt}
I.~F. Ginzburg and I.~P. Ivanov, {\it {Tree-level unitarity constraints in the
  most general 2HDM}},  {\em Phys. Rev.} {\bf D72} (2005) 115010,
  [\href{http://xxx.lanl.gov/abs/hep-ph/0508020}{{\tt hep-ph/0508020}}].

\bibitem{Grinstein:2015rtl}
B.~Grinstein, C.~W. Murphy, and P.~Uttayarat, {\it {One-loop corrections to the
  perturbative unitarity bounds in the CP-conserving two-Higgs doublet model
  with a softly broken $ {\mathrm{\mathbb{Z}}}_2 $ symmetry}},  {\em JHEP} {\bf
  06} (2016) 070, [\href{http://xxx.lanl.gov/abs/1512.04567}{{\tt
  1512.04567}}].

\bibitem{Cacchio:2016qyh}
V.~Cacchio, D.~Chowdhury, O.~Eberhardt, and C.~W. Murphy, {\it {Next-to-leading
  order unitarity fits in Two-Higgs-Doublet models with soft $\mathbb{Z}_2$
  breaking}},  {\em JHEP} {\bf 11} (2016) 026,
  [\href{http://xxx.lanl.gov/abs/1609.01290}{{\tt 1609.01290}}].

\bibitem{Blankenburg:2012ex}
G.~Blankenburg, J.~Ellis, and G.~Isidori, {\it {Flavour-Changing Decays of a
  125 GeV Higgs-like Particle}},  {\em Phys. Lett.} {\bf B712} (2012) 386--390,
  [\href{http://xxx.lanl.gov/abs/1202.5704}{{\tt 1202.5704}}].

\bibitem{Aad:2014dya}
{\bf ATLAS} Collaboration, G.~Aad {\em et~al.}, {\it {Search for top quark
  decays $t \to qH$ with $H \to \gamma\gamma$ using the ATLAS detector}},  {\em
  JHEP} {\bf 06} (2014) 008, [\href{http://xxx.lanl.gov/abs/1403.6293}{{\tt
  1403.6293}}].

\bibitem{Khachatryan:2014jya}
{\bf CMS} Collaboration, V.~Khachatryan {\em et~al.}, {\it {Searches for heavy
  Higgs bosons in two-Higgs-doublet models and for $t\to ch$ decay using
  multilepton and diphoton final states in $pp$ collisions at 8 TeV}},  {\em
  Phys. Rev.} {\bf D90} (2014) 112013,
  [\href{http://xxx.lanl.gov/abs/1410.2751}{{\tt 1410.2751}}].

\bibitem{Khachatryan:2016atv}
{\bf CMS} Collaboration, V.~Khachatryan {\em et~al.}, {\it {Search for top
  quark decays via Higgs-boson-mediated flavor-changing neutral currents in pp
  collisions at $\sqrt{s} =$ 8 TeV}},
  \href{http://xxx.lanl.gov/abs/1610.04857}{{\tt 1610.04857}}.

\bibitem{Botella:2012ab}
F.~Botella, G.~Branco, and M.~Rebelo, {\it {Invariants and Flavour in the
  General Two-Higgs Doublet Model}},  {\em Phys.Lett.} {\bf B722} (2013)
  76--82, [\href{http://xxx.lanl.gov/abs/1210.8163}{{\tt 1210.8163}}].

\bibitem{Bernabeu:1986fc}
J.~Bernabeu, G.~C. Branco, and M.~Gronau, {\it {CP restrictions on quark mass
  matrices}},  {\em Phys. Lett.} {\bf 169B} (1986) 243--247.

\bibitem{Davidson:2005cw}
S.~Davidson and H.~E. Haber, {\it {Basis-independent methods for the
  two-Higgs-doublet model}},  {\em Phys. Rev.} {\bf D72} (2005) 035004,
  [\href{http://xxx.lanl.gov/abs/hep-ph/0504050}{{\tt hep-ph/0504050}}].
  [Erratum: Phys. Rev.D72,099902(2005)].

\bibitem{Wolfenstein:1983yz}
L.~Wolfenstein, {\it {Parametrization of the Kobayashi-Maskawa Matrix}},  {\em
  Phys.Rev.Lett.} {\bf 51} (1983) 1945.

\bibitem{Trodden:1998ym}
M.~Trodden, {\it {Electroweak baryogenesis}},  {\em Rev. Mod. Phys.} {\bf 71}
  (1999) 1463--1500, [\href{http://xxx.lanl.gov/abs/hep-ph/9803479}{{\tt
  hep-ph/9803479}}].

\bibitem{Morrissey:2012db}
D.~E. Morrissey and M.~J. Ramsey-Musolf, {\it {Electroweak baryogenesis}},
  {\em New J. Phys.} {\bf 14} (2012) 125003,
  [\href{http://xxx.lanl.gov/abs/1206.2942}{{\tt 1206.2942}}].

\bibitem{Jarlskog:1985ht}
C.~Jarlskog, {\it {Commutator of the Quark Mass Matrices in the Standard
  Electroweak Model and a Measure of Maximal CP Violation}},  {\em Phys. Rev.
  Lett.} {\bf 55} (1985) 1039.

\bibitem{Branco:1999fs}
G.~C. Branco, L.~Lavoura, and J.~P. Silva, {\em {CP Violation}, International
  Series of Monographs on Physics, Oxford University Press}.

\bibitem{Nebot:2015wsa}
M.~Nebot and J.~P. Silva, {\it {Self-cancellation of a scalar in neutral meson
  mixing and implications for the LHC}},  {\em Phys. Rev.} {\bf D92} (2015),
  no.~8 085010, [\href{http://xxx.lanl.gov/abs/1507.07941}{{\tt 1507.07941}}].

\bibitem{Barr:1990vd}
S.~M. Barr and A.~Zee, {\it {Electric Dipole Moment of the Electron and of the
  Neutron}},  {\em Phys. Rev. Lett.} {\bf 65} (1990) 21--24. [Erratum: Phys.
  Rev. Lett.65,2920(1990)].

\bibitem{Weinberg:1989dx}
S.~Weinberg, {\it {Larger Higgs Exchange Terms in the Neutron Electric Dipole
  Moment}},  {\em Phys. Rev. Lett.} {\bf 63} (1989) 2333.

\bibitem{Jung:2013hka}
M.~Jung and A.~Pich, {\it {Electric Dipole Moments in Two-Higgs-Doublet
  Models}},  {\em JHEP} {\bf 04} (2014) 076,
  [\href{http://xxx.lanl.gov/abs/1308.6283}{{\tt 1308.6283}}].

\bibitem{BowserChao:1997bb}
D.~Bowser-Chao, D.~Chang, and W.-Y. Keung, {\it {Electron electric dipole
  moment from CP violation in the charged Higgs sector}},  {\em Phys. Rev.
  Lett.} {\bf 79} (1997) 1988--1991,
  [\href{http://xxx.lanl.gov/abs/hep-ph/9703435}{{\tt hep-ph/9703435}}].

\bibitem{Ilisie:2015tra}
V.~Ilisie, {\it {New Barr-Zee contributions to $\mathbf{(g-2)_\mu}$ in
  two-Higgs-doublet models}},  {\em JHEP} {\bf 04} (2015) 077,
  [\href{http://xxx.lanl.gov/abs/1502.04199}{{\tt 1502.04199}}].

\end{thebibliography}

\providecommand{\href}[2]{#2}\begingroup\raggedright\endgroup

\end{document}